\pdfoutput=1 
\documentclass[a4paper, 11pt]{article}
\usepackage[dvipsnames]{xcolor}

\usepackage{jheppub} 

\makeatletter  
\gdef\@fpheader{}  
\makeatother 

\usepackage{bm}
\usepackage{braket}

\usepackage{blindtext}
\usepackage{epsfig, cancel}

\usepackage{latexsym}
\usepackage{natbib, comment}
\usepackage{mathrsfs,amsmath,amssymb,amsthm,amsfonts,tikz,graphicx,accents,hyperref,color}
\usepackage{url}
\usepackage{dcolumn}
\usepackage{multirow}
\usepackage{color}
\usepackage{cancel}
\usepackage{soul}
\usepackage[normalem]{ulem}
\usepackage{txfonts}
\usepackage{epsfig}
\usepackage{psfrag}
\usepackage{subfigure}
\hypersetup{colorlinks=true}
\usepackage{mathtools}
\usepackage{enumitem}
\usepackage{float}
\usetikzlibrary{intersections,calc}

\def\H0{{\text{H}\hspace*{-2.05mm}\text{H} 0\hspace*{-1.35mm}0\ }}

\usepackage[all]{xy}

\usepackage{amssymb}
\usepackage{ragged2e}
\usepackage{slashed}
\usepackage{slashed,ccaption}
\usepackage{multirow}

\usepackage{caption}

\usepackage{array}
\hypersetup{ linktoc=all,
    colorlinks, linkcolor={palatinateblue},
    citecolor={brightpink}, urlcolor={amaranth}
}

\graphicspath{{Images/}}

\renewcommand{\d}[1]{\ensuremath{\operatorname{d}\!{#1}}}

\DeclareSymbolFont{extraup}{U}{zavm}{m}{n}
\DeclareMathSymbol{\varheart}{\mathalpha}{extraup}{86}
\DeclareMathSymbol{\vardiamond}{\mathalpha}{extraup}{87}
\makeatletter
\renewcommand*{\@fnsymbol}[1]{\ensuremath{\ifcase#1\or \clubsuit \or \vardiamond \or \varheart\or
    \spadesuit\or \mathparagraph\or \|\or **\or \dagger\dagger
    \or \ddagger\ddagger \else\@ctrerr\fi}}
\makeatother

\definecolor{rosy}{RGB}{230,235,252}
\definecolor{myframetitle}{RGB}{90,89,170}
\definecolor{myblocktitle}{RGB}{140,185,249}
\definecolor{mytitle}{RGB}{10,80,26}

\definecolor{darkgreen}{RGB}{27,130,45}
\definecolor{darkblue}{rgb}{0,0,0.3}
\definecolor{darkred}{rgb}{0.7,0,0}

\definecolor{light gray}{RGB}{220,220,220}
\definecolor{dark purple}{RGB}{108,0,217}
\definecolor{pink}{RGB}{190,20,100}
\definecolor{orang}{RGB}{193,63,0}
\definecolor{green}{RGB}{11,98,17}
\definecolor{darkpink}{RGB}{153,0,76}
\definecolor{bluegreen}{RGB}{0,102,102}
\definecolor{greenlagan}{RGB}{0,102,0}
\definecolor{redgreen}{RGB}{102,102,0}
\definecolor{Redgreen}{RGB}{153,76,0}
\definecolor{vividviolet}{rgb}{0.62, 0.0, 1.0}
\definecolor{amaranth}{rgb}{0.9, 0.17, 0.31}
\definecolor{palatinateblue}{rgb}{0.15, 0.23, 0.89}
\definecolor{brightpink}{rgb}{1.0, 0.0, 0.5}
\definecolor{cornflowerblue}{rgb}{0.39, 0.58, 0.93}
\definecolor{deepcarminepink}{rgb}{0.94, 0.19, 0.22}
\definecolor{radicalred}{rgb}{1.0, 0.21, 0.37}

\DeclareFontFamily{OT1}{rsfs}{}

\DeclareFontShape{OT1}{rsfs}{m}{n}{ <-7> rsfs5 <7-10> rsfs7 <10->rsfs10}{} 

\DeclareMathAlphabet{\mycal}{OT1}{rsfs}{m}{n}

\newcommand{\be}{\begin{equation}}
\newcommand{\ee}{\end{equation}}

\usepackage[most]{tcolorbox}

\tcbset{highlight math style={left=02mm,right=02mm,top=02mm,bottom=02mm}} 
\usepackage{empheq}

\newcommand\inbox[1]{\tcbset{fonttitle=\scriptsize} \tcboxmath[colback=white,colframe=black!70]{#1}}

\newcommand{\Ottbar}{{\mathcal{T}\hspace*{-1mm}{\mathcal{T}}}}

\newcommand{\ttbarb}{\mathrm{T}\hspace*{-1mm}\mathrm{T}}

\begin{document}

\title{AdS$_3$ Freelance Holography, A Detailed Analysis}

\author[a]{M.M.~Sheikh-Jabbari}
\author[a,b]{, V.~Taghiloo}

\affiliation{$^a$ School of Physics, Institute for Research in Fundamental
Sciences (IPM),\\ P.O.Box 19395-5531, Tehran, Iran}
\affiliation{$^b$ Department of Physics, Institute for Advanced Studies in Basic Sciences (IASBS),\\ 
P.O. Box 45137-66731, Zanjan, Iran}
\emailAdd{
jabbari@theory.ipm.ac.ir, v.taghiloo@iasbs.ac.ir}
\abstract{Freelance holography program is an extension of the gauge/gravity correspondence in which the boundary theory can reside on any timelike codimension-one surface in AdS space, and the boundary conditions on the bulk fields can be chosen arbitrarily. Freelance holography provides the framework for a systematic study of various boundary conditions and associated bulk geometries. In this work, we analyze in detail the AdS$_3$ freelance holography.  
One can explicitly solve for the bulk  AdS$_3$ Einstein gravity equations of motion. For generic boundary conditions, the solutions are described by two arbitrary functions of one variable. We study holographic renormalization group (RG) flows, the interpolation between different boundary conditions at different boundaries and associated surface charges and their algebras.
}
\maketitle

\section{Introduction}\label{sec:Introduction}

The AdS$_{d+1}$/CFT$_d$ establishes a duality between a $d$-dimensional quantum field theory (QFT), typically a gauge field theory, and a $(d+1)$-dimensional (quantum) gravity on (asymptotically) AdS$_{d+1}$ spacetime \cite{Aharony:1999ti, Witten:1998zw, Gubser:1998bc}. In the duality, the QFT resides on the boundary of the AdS space, and the gravity resides in the bulk. So, the QFT, which is generically a non-gravitating theory, is referred to as ``boundary theory'' and the gravity theory the ``bulk'' theory.  Gauge/gravity correspondence is a limit of the AdS/CFT duality, where the gravity side is well captured by the classical Einstein gravity (plus higher curvature corrections) and the QFT side is reduced to the planar sector described by the large $N$ limit \cite{tHooft:1973alw} ($N$ being the rank of the gauge group of the QFT). 

To define the gravity theory (the bulk side), one needs to specify the boundary conditions of the fields on the boundary of the AdS space. The standard AdS/CFT one is prescribed to adopt Dirichlet boundary conditions on the AdS asymptotic causal/conformal boundary of the AdS space \cite{Aharony:1999ti, Witten:1998zw, Henningson:1998gx}. This boundary condition is guaranteed through adding the Gibbons-Hawking-York boundary term \cite{Gibbons:1976ue, York:1972sj} and is a convenient choice, because Dirichlet boundary conditions freeze the fluctuations of the metric at the boundary, the metric on the spacetime in which the boundary QFT resides \cite{Balasubramanian:1999re, Henningson:1998gx}, dovitailing with the fact that the boundary theory is a non-gravitating QFT on a given spacetime. 

The AdS$_{d+1}$ is conveniently foliated by the $ d$-dimensional timelike part and the ``holographic'' direction/the radial direction. In the AdS/CFT setup, the holographic direction plays the role of energy scale, renormalization scale, in which the dual QFT is defined \cite{deBoer:1999tgo, Verlinde:1999xm, Bianchi:2001kw, Akhmedov:2010sw}. In the well-established Wilsonian EFT, one can define any QFT at energy scales below any given energy scale $\Lambda$ at any desired precision by specifying the coupling of the theory at $\Lambda$ and then applying the renormalization group (RG) and coarse-graining/integrating out, procedure \cite{Wilson:1973jj, Polchinski:1983gv, Georgi:1993mps, Burgess:2007pt}. In particular, in this prescription, one need not start from extreme UV to define the theory. Therefore, to match up with the Wilsonian picture, one needs to extend the AdS/CFT or gauge/gravity correspondence to the cases where the AdS space is cutoff at an arbitrary scale/radius (and the AdS space is not extended all the way to its asymptotic boundary), see Fig.\ref{fig:ADS-timelike}.

To define the AdS/CFT on a cutoff AdS space, one should specify the boundary condition, and unlike the asymptotic case, where the Dirichlet boundary condition had a clear physical motivation, at a finite cutoff, other choices may be relevant for holographic descriptions. From a different viewpoint, it was noted that multi-trace deformations on the QFT side correspond to the addition of boundary terms to the gravity side \cite{Witten:2001ua, Berkooz:2002ug, Sever:2002fk}. As stressed in \cite{Parvizi:2025shq}, these boundary terms modify the Dirichlet boundary conditions. Besides multi-trace deformations, other boundary conditions on the asymptotic boundary of AdS space have also been studied, with or without a holographic picture and applications in mind \cite{Compere:2008us, Ishibashi:2004wx, Marolf:2006nd}. 

With the above motivations, in references \cite{Parvizi:2025shq, Parvizi:2025wsg} (see \cite{Taghiloo:2025oeu} for a non-technical review), we started developing the ``freelance holography'' program: a systematic formulation for AdS/CFT (more precisely gauge/gravity correspondence) in which boundary (where the QFT resides) as well as the boundary conditions on the bulk fields can be chosen arbitrarily. That is, we set the boundary and boundary conditions free in a holographic setting. As a very intriguing and quite non-trivial outcome, in \cite{Adami:2025pqr} we discussed that the freelance program leads us to the fact that gravity is an effective description of certain deformations by the square of the energy-momentum tensor arising in RG flow (see also \cite{Ran:2025xas, Li:2025lpa} for related recent works).\footnote{{See also \cite{Dubovsky:2017cnj, Dubovsky:2018bmo, Cardy:2018sdv, Tolley:2019nmm, Hirano:2025cjg, Hirano:2025tkq} for alternative perspectives on the connection between two-dimensional T$\bar{\text{T}}$ deformations and the corresponding two-dimensional gravity theories.}} This work is a step forward in developing further the freelance holography program.

In this work, we focus on the AdS$_3$/CFT$_2$ example and make a detailed and explicit analysis of the general freelance holography framework discussed in \cite{Parvizi:2025shq, Parvizi:2025wsg}. AdS$_3$ Einstein gravity has had a pivotal role in the development of AdS/CFT, as the seminal Brown-Henneaux analysis \cite{Brown:1986nw} has been viewed as a precursor to AdS/CFT. Moreover, AdS$_3$ gravity does not have propagating degrees of freedom and its equations can be explicitly integrated; it is simple enough to be handled explicitly, while capturing the key features needed for a nice demonstration of the freelance holography program. We systematically analyze several boundary conditions for AdS$_3$ gravity and construct the associated solution phase space associated with each boundary condition. While we recover all previous cases discussed in previous literature \cite{Banados:1998gg, Afshar:2016wfy, Afshar:2016kjj, Afshar:2016uax, Afshar:2017okz, Adami:2020uwd, Compere:2013bya}, we construct new solutions. Besides providing a unified description of all AdS$_3$ solutions in the literature, we discuss the physics of the new solutions. 
\paragraph{Organization of the paper.}
In Section~\ref{sec:Basic-setup}, we present the essential ingredients for our analysis, including the integration of the AdS$_3$ gravity equations of motion and a brief review of the freelance holography framework. Section~\ref{sec:classification-bc} provides a systematic way to specify different types of boundary conditions in AdS$_3$ gravity, with emphasis on the constraints that these conditions must satisfy. In Sections~\ref{sec: bdry-constraints} and \ref{sec:noncovariant-bc-solnspace}, we solve the constraint equations under various boundary conditions at the AdS$_3$ asymptotic boundary and construct various class of solutions. While some of these solutions have been previously discussed in the literature, we introduce and study some different classes of new solutions. Section~\ref{sec:bc-finite} focuses on solutions with various boundary conditions imposed at a finite radial cutoff. Section~\ref{sec:discussion} concludes the paper with a summary of our findings and a discussion of future directions. In Appendix~\ref{appen:useful-rel}, we collect useful relations for the radial dependence of various quantities, which interpolate between geometric data at finite distance and their asymptotic counterparts. Appendix~\ref{appen:non-flat-bc} is devoted to solutions with different boundary conditions on a non-flat AdS boundary. In Appendix~\ref{appen:charges}, we present analysis of surface charges associated with different classes of solutions discussed in sections \ref{sec: bdry-constraints} and \ref{sec:noncovariant-bc-solnspace}.

\section{Preliminaries}\label{sec:Basic-setup}
In this section, we present the geometric setup that forms the basis for our subsequent analysis. We then focus on three-dimensional Einstein gravity with a negative cosmological constant, where we solve the field equations and determine the exact radial dependence of the line element.

\subsection{\texorpdfstring{AdS$_3$}{AdS3} gravity }\label{sec:ADS-gravity}

\begin{figure}[t]
\centering
\begin{tikzpicture}[scale=1.2]

\node (v1) at (0.8,0) {};
\node (v4) at (0.8,-3) {};
\node (v5) at (-0.8,0) {};
\node (v8) at (-0.8,-3) {};

\begin{scope}
  \clip ($(v1)+(0,0)$) to[out=135,in=45, looseness=0.5] ($(v5)+(0,0)$)
        -- ($(v8)+(0,0)$) to[out=45,in=135, looseness=0.5] ($(v4)+(0,0)$)
        -- cycle;
  \fill[gray!20] (-0.8,0) -- (-0.8,-3) -- (0.8,-3) -- (0.8,0) -- cycle;
\end{scope}

\draw [darkred!60, very thick] (-0.8,0) -- (-0.8,-3);
\draw [darkred!60, very thick] (0.8,0) -- (0.8,-3);

\begin{scope}[fill opacity=0.8, very thick, darkred!60]
  \filldraw [fill=gray!30] (0.8,0) 
    to[out=135,in=45, looseness=0.5] (-0.8,0)
    to[out=315,in=225,looseness=0.5] (0.8,0);
\end{scope}
\begin{scope}[fill opacity=0.8, very thick, darkred!60]
  \filldraw [fill=gray!30] (0.8,-3) 
    to[out=135,in=45, looseness=0.5] (-0.8,-3)
    to[out=315,in=225,looseness=0.5] (0.8,-3);
\end{scope}

\draw [blue!60](0,0) ellipse (1.5 and 0.5);
\draw [blue!60] (0,-3) ellipse (1.5 and 0.5);
\draw [blue!60] (-1.5,0) -- (-1.5,-3);
\draw [blue!60] (1.5,0) -- (1.5,-3);

\fill (0,-1.5)  node [blue!50!black] {${\cal M}_r$};
\fill (0.8,-2)+(0.3,0.5) node [darkred!60] {$\Sigma_r$};
\fill (0.8,-3)+(0.7,0.8) node [right,blue!80] {$\Sigma$};
\fill (0, -3.8) node [black] { $\mathcal{M}$};
\end{tikzpicture}
\caption{ \justifying{\footnotesize{ Asymptotically AdS$_3$ spacetime and a region cutoff at radius $r$.  $\textcolor{black}{\cal{M}}$ denotes the global asymptotically AdS$_3$ spacetime and  $\textcolor{blue!80}{\Sigma}$ is its asymptotic timelike boundary.  The shaded region $\textcolor{blue!50!black}{\mathcal{M}_r}$ is the part of AdS$_3$ cutoff at radius $r$, enclosed in a timelike surface $\textcolor{darkred!60}{\Sigma_r}$.}}}
\label{fig:ADS-timelike}
\end{figure}
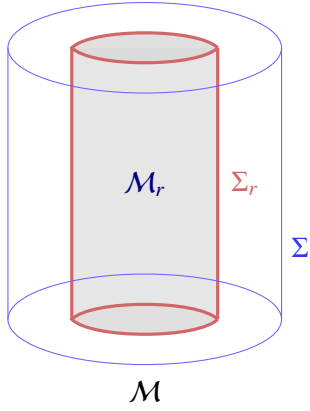

\paragraph{Radial decomposition.} Consider a \(3\)-dimensional asymptotically AdS spacetime \(\mathcal{M}\) with coordinates \(x^\mu\) and metric \(g_{\mu\nu}\). We foliate  \(\mathcal{M}\) by a family of codimension-one timelike hypersurfaces \(\Sigma_r\), labeled by the radial coordinate \(r \in (r_\circ, \infty)\), where $r_\circ \geq 0$ and  decompose the spacetime coordinates as \(x^\mu = (x^a, r)\), where \(a = 0,1\). \(\Sigma_r\) are constant-\(r\) timelike slices spanned by \(x^a\)  coordinates. We assume that the AdS boundary lies at \(r = \infty\) and refer to the asymptotic timelike boundary as \(\Sigma:=\Sigma_{r=\infty}\). For a finite cutoff at \(r\), we define \(\mathcal{M}_r\) as the portion of AdS spacetime bounded to \(\Sigma_r\); see Figure \ref{fig:ADS-timelike}. 

Using this foliation, we can perform a radial \((1+2)\)-dimensional ADM decomposition of the metric where the line element then takes the standard form  
\begin{equation}\label{metric}
   \d s^2 = N^2 \d r^2 + h_{ab} (\d x^{a} + U^{a} \d r)(\d x^{b} + U^{b} \d r)\, ,
\end{equation}  
where \(N\) denotes the radial lapse function, \(U^a\) the radial shift vector, and \(h_{ab}\) the induced metric on the constant-\(r\) hypersurface \(\Sigma_r\). For convenience, we introduce a conformally rescaled metric, $ h_{ab} = r^2 \gamma_{ab}.$
At this stage, no specific gauge choice has been imposed; both \(N\) and \(U^a\) remain free functions that can be fixed as needed. This setup can be regarded as a generalized Fefferman--Graham (FG) expansion \cite{Fefferman:1985abc}. The normal one-form to \(\Sigma_r\), together with the corresponding induced metric, are then given by
\begin{equation}
    s = s_{\mu} \d x^{\mu} = N \d r \, , \qquad h_{\mu\nu} = g_{\mu\nu} - s_{\mu}s_{\nu} \, .
\end{equation}  
For later use, we define the extrinsic curvature of \(\Sigma_r\) as  
\begin{equation}\label{K-form}
    K_{\alpha \beta}=\frac{1}{2} h^{\mu}{}_{\alpha}\, h^{\nu}_{\beta}\, \mathcal{L}_{s} h_{\mu\nu}\, \quad \Longrightarrow \quad   K_{ab} = \frac{1}{2N} \mathcal{D}_{r} h_{ab} \, ,
\end{equation}  
where $\mathcal{D}_{r} := \partial_{r} - \mathcal{L}_{U}$ is an operator and \(\mathcal{L}_{U}\) stands for the Lie derivative along shift vector \(U^a\).  

In our analysis here, and to keep equations more tractable, we will work in the FG gauge in which, 
\begin{equation}\label{FG-gauge}
    N=\frac{\ell}{r}, \qquad U^a=0,
\end{equation}
where $\ell$ is the AdS$_3$ radius. 

\paragraph{AdS$_3$ Einstein gravity action.}\label{sec:AdS3-gravity}

The AdS$_3$ Einstein-Hilbert action defined on a portion $\mathcal{M}_r$ of an asymptotically $\text{AdS}_3$ spacetime, with Dirichlet boundary conditions  on its radial boundary $\Sigma_r$
\begin{equation}\label{action-AdS-3}
     \hspace{-.08 cm} S^{\text{\tiny{D}}}_{\text{\tiny{bulk}}}[\mathcal{M}_r] = \frac{1}{2\kappa} \int_{\mathcal{M}_r} \sqrt{-g} \left(\mathscr{R} + \frac{2}{\ell^2}\right) + \frac{1}{\kappa} \int_{\Sigma_r} \sqrt{-h}\,  (K - \ell^{-1})\, , 
\end{equation} 
where \( \kappa = 8\pi G \), \( \ell \) is the AdS\(_{3}\) radius and $\mathscr{R}$ denotes the scalar curvature of bulk metric $g_{\mu\nu}$. The boundary term in \eqref{action-AdS-3} consists of the Gibbons-Hawking-York (GHY) term along with counterterms \cite{Balasubramanian:1999re, Emparan:1999pm}.

Varying the action \eqref{action-AdS-3} yields the Einstein field equations,  
\begin{equation}\label{eom-AdS3}
    \mathscr{R}_{\mu\nu}-\frac{1}{2}\mathscr{R}\, g_{\mu\nu} -\frac{1}{\ell^2} g_{\mu\nu}=0\, , \qquad \mathscr{R}= -\frac{6}{\ell^2}\, ,
\end{equation}  
and the Dirichlet symplectic potential,  
\begin{equation}\label{sym-pot-r}
    \Theta_{\text{\tiny{D}}}[\Sigma_r] := \delta S^{\text{\tiny{D}}}_{\text{\tiny{bulk}}}[\mathcal{M}_r]\big|_{\text{on-shell}}=-\frac{1}{2} \int_{\Sigma_r} \sqrt{-h}\, \mathcal{T}^{ab}\, \delta h_{ab} +\text{corner terms}\, ,
\end{equation}  
where \( \mathcal{T}^{ab} \) is the renormalized Brown-York energy-momentum tensor (rBY-EMT) \cite{Balasubramanian:1999re} \footnote{The standard Brown–York energy–momentum tensor \cite{Brown:1992br} is defined by \eqref{BY-EMT-unren}, excluding the last term.}
\begin{equation}\label{BY-EMT-unren}
     {\mathcal{T}}^{ab}=\frac{1}{\kappa}\left(K^{ab}-K h^{ab}+\ell^{-1} h^{ab}\right)\, .
\end{equation}
The corner terms in \eqref{sym-pot-r} are those involving integration over codimension 2 surfaces, boundary of $\Sigma_r$ (e.g., constant time slices over $\Sigma_r$). These terms and their role in the freelance holography program have been discussed in detail in \cite{Parvizi:2025shq, Parvizi:2025wsg}. In this work, unless explicitly discussed and mentioned, we generically drop such terms.

Einstein equations \eqref{eom-AdS3} in the FG gauge \eqref{FG-gauge} take the form
\begin{subequations}\label{EoM-GR-mat-decompose}
    \begin{align}
       & \mathcal{E}_{rr} := \mathcal{T} + {\frac{{c} }{24\pi}} {\cal R} + \frac{6\pi\ell^2}{{c}}\Ottbar  = 0 \, ,\label{EoM-ss}\\
       & \mathcal{E}_{ra} := {\nabla}_b{\mathcal{T}}^{b}_{a}=0\, , \label{EoM-sa}\\
       & \mathcal{E}_{ab} := r\partial_{r} \mathcal{T}_{ab} - \frac{12\pi \ell^2}{{c}}\left(\mathcal{T}\mathcal{T}_{ab}+\frac{3}{2} \Ottbar h_{ab}\right)=0\, , \label{EoM-ab}
    \end{align}
\end{subequations}
where $c:=\frac{12\pi \ell}{\kappa}$ is the Brown-Henneaux central charge \cite{Brown:1986nw}, \( {\cal R} \) is the Ricci scalar of metric \( h_{ab} \) on \( \Sigma_{r} \), $\nabla_a$ is the covariant derivative associated with  metric $h_{ab}$ and $\Ottbar$ is the T$\bar{\text{T}}$ operator \cite{Zamolodchikov:2004ce, Smirnov:2016lqw}
\begin{equation}\label{ottbar}
   \Ottbar := {\mathcal{T}}^{ab}\,{\mathcal{T}}_{ab} - {\mathcal{T}}^2\, .
\end{equation} 
\paragraph{Integrating field equations.}\label{sec:integrability}
The radial dependence of the equations of motion exactly—a feature often referred to as \textit{integrability}.  To begin with, recall that the Bianchi identities and their contracted form imply  
\begin{equation}
  \partial_r\!\left( \sqrt{-h}\, \mathcal{E}_{ra} \right) = 0 \,, 
  \qquad 
  \partial_r\!\left( \sqrt{-h}\, \mathcal{E}_{rr} \right) = 0 \,.  
\end{equation}  
When viewed as evolution equations in the radial direction, the conditions \(\mathcal{E}_{rr}=0\) and \(\mathcal{E}_{ra}=0\) respectively represent the Hamiltonian and momentum constraints, which must hold on every constant-\(r\) slice. In particular, if these constraints are satisfied asymptotically—at large \(r\) near the AdS\(_3\) causal boundary—then any solution of \eqref{EoM-ab} will automatically satisfy them at arbitrary \(r\).  

We now turn to the explicit solution of the radial dependence. In the Fefferman--Graham gauge, \eqref{K-form} and \eqref{BY-EMT-unren} can be written as
\begin{equation}\label{hab-r-T}
\begin{split}
   r\partial_{r} {h}_{ab}= 2h_{ab}+ \frac{24\pi \ell^2}{c} \tilde{\mathcal{T}}_{ab}\, ,
\end{split}
\end{equation}
where 
\begin{equation}
    \tilde{\mathcal{T}}_{ab}:={\mathcal{T}}_{ab}-{\cal T} h_{ab}\, ,
\end{equation}
is the trace-reversed rBY-EMT. 
So, \eqref{EoM-ab} is a second-order differential equation in $r$ for $h_{ab}$. This equation can be integrated in the $3d$ case and one can completely determine the $r$ dependence of $h_{ab}$ and hence ${\cal T}_{ab}$ in terms of two ``integration constants''. Using the $2d$ identities,
\begin{equation}
    {\cal R}_{ab}=\frac12 {\cal R} h_{ab}\, , \qquad 2\mathcal{T}_{ac}\mathcal{T}^{c}_{b}=2\mathcal{T}\, \mathcal{T}_{ab}+\Ottbar\, h_{ab}\, , \qquad 2\tilde{\mathcal{T}}_{ac}\mathcal{T}^{c}_{b}=\Ottbar\, h_{ab}\,,
\end{equation}
and after some straightforward algebra, we obtain, 
\begin{equation}\label{h-coord-trans}
\begin{split}
 \inbox{ h_{ab}=\left(\frac{r}{\ell}q_{ac} - \frac{6\pi\ell^3}{c r}\tilde{T}_{ac}\right) q^{cd}\left(\frac{r}{\ell}q_{bd} - \frac{6\pi\ell^3}{c r} \tilde{T}_{bd}\right)\, . }
\end{split}\end{equation}
Here $q^{ab}$ is inverse of $q_{ab}$, $q_{ab}q^{bc}=\delta_a^c$ and 
\begin{equation}
    \begin{split}
        \tilde{T}_{ab} &:= T_{ab} -T q_{ab}\, ,
    \end{split}
\end{equation}
is the trace-reversed EMT at infinity. 
Therefore, \eqref{EoM-ab} determines the induced metric up to a codimension-one tensor $q_{ab}$ and the stress tensor $T_{ab}$. These integration constants are still subject to Hamiltonian and momentum constraints \eqref{EoM-ss} and \eqref{EoM-sa}, i.e.
\begin{equation}\label{cons-trace}
    D_{a}T^{ab}=0\, ,  \quad   \quad T:=q^{ab}\, T_{ab}=-\frac{c}{24 \pi} {R}\, ,
\end{equation}
where \(D_a\) and \(R\) denote the covariant derivative and Ricci scalar associated with \(q_{ab}\), and the indices of \(T_{ab}\) are raised using the inverse metric \(q^{ab}\). The second equation in \eqref{cons-trace} corresponds to the well-known trace anomaly equation \cite{Henningson:1998gx}.  

Finally, we note that \(q_{ab}\) and \(T^{ab}\) represent the leading contributions (at large \(r\)) to the boundary metric \(h_{ab}\) and the rBY-EMT \({\cal T}_{ab}\), respectively
\begin{equation}
    \begin{split}
     q_{ab} = \lim_{r \to \infty} \frac{\ell^2}{r^2}\, h_{ab} \, ,\qquad 
       \tilde{T}_{ab} = \lim_{r \to \infty} \tilde{\mathcal{T}}_{ab}\, .
    \end{split}
\end{equation}

Thus far, we have an explicit solution of the $r$ dependence of all relevant quantities, matching known results in the literature \cite{McGough:2016lol, Guica:2019nzm, AliAhmad:2025kki}, see appendix \ref{appen:useful-rel} for more explicit relations among different quantities at $r$ and at $r\to\infty$. These identities and relations will be used in constructing explicit solutions with generic prescribed boundary conditions. These generic solutions are  determined by two covariantly constant symmetric tensors $q_{ab}$ and $T^{ab}$. Therefore, the most general solution is specified by 4 unconstrained boundary data. Of course, one should still impose boundary conditions and \eqref{EoM-ss} on these 4 functions. We shall do so in the following sections.

\subsection{\texorpdfstring{$3d$}{3d} freelance holography}\label{sec:freelance-section}

In this section, we review the freelance holography framework \cite{Parvizi:2025shq, Parvizi:2025wsg, Taghiloo:2025oeu} customized to the $3d$ case. This setup allows for different boundary conditions on the metric field $h_{ab}$ and/or ${\cal T}_{ab}$ at an arbitrary boundary $\Sigma_r$. 

\subsubsection{Brief review on covariant phase space formalism}\label{sec:CPSF}
The freelance holography can be conveniently formulated within the Covariant Phase Space Formalism  (CPSF), first developed in  \cite{Lee:1990nz, Iyer:1994ys, Wald:1999wa} (see also \cite{Grumiller:2022qhx, Parvizi:2025shq,  Parvizi:2025wsg}). Consider an action in $\mathcal{M}_r$ region \eqref{bulk-action-W} with boundary $\Sigma_r$ 
\begin{equation}\label{bulk-action-W}
S^{\text{\tiny{W}}}_{\text{\tiny{bulk}}}[\mathcal{M}_r] = S^{\text{\tiny{D}}}_{\text{\tiny{bulk}}}[\mathcal{M}_r] + \int_{\Sigma_r} n_{\mu} W^{\mu}\, ,
\end{equation}
where $n = n_{\mu} \, \mathrm{d}x^{\mu} = \mathrm{d}r$ denotes the unnormalized normal one-form to the boundary $\Sigma_r$, and $S^{\text{\tiny D}}_{\text{\tiny bulk}}[\mathcal{M}_r]$ is the action subject to Dirichlet boundary conditions on $\Sigma_r$ \eqref{action-AdS-3}. The role of the $W$-term is to accommodate the desired (arbitrary) boundary conditions. To see this, we work out the  on-shell variation of \eqref{bulk-action-W},
\begin{equation}\label{symp-pot-W}
\Theta^{\text{\tiny{W}}}[\Sigma_r] :=\delta S^{\text{\tiny{W}}}_{\text{\tiny{bulk}}}[\mathcal{M}_r]\big|_{\text{on-shell}}= \Theta^{\text{\tiny{D}}}[\Sigma_r] + \delta\int_{\Sigma_r} n_{\mu} W^{\mu}\, ,
\end{equation}
where \(\Theta^{\text{\tiny D}}[\Sigma_r]\) denotes the Dirichlet symplectic potential \eqref{sym-pot-r}, consistent with the Dirichlet boundary condition. In this case, the Dirichlet solution space is labeled by the pair \(\{h_{ab}, \mathcal{T}^{ab}\}\) \eqref{sym-pot-r}, and the symplectic potential vanishes when \(\delta h_{ab}\big|_{\Sigma_r}=0\). In the presence of an additional boundary term \(W\), one moves from the Dirichlet labels to a new pair, which enforces a modified boundary condition. The explicit form of this pair, and the corresponding boundary condition, is determined by the \(W\) term.

Finally, we note that the definition of the symplectic potential is not uniquely determined by a given action and boundary condition; there exist codimension-two freedoms, which can be expressed as follows \cite{Iyer:1994ys, Wald:1999wa} (see also \cite{Parvizi:2025shq})
\begin{equation}\label{Y-Z-freedom}
    \tilde{\Theta}^{\text{\tiny{W}}}[\Sigma_r] = \Theta^{\text{\tiny{W}}}[\Sigma_r] + \int_{\Sigma_{r}}\, n_{\mu} (\partial_{\nu} Y^{\mu\nu} + \partial_{\nu} \delta Z^{\mu\nu}  )\, ,
\end{equation}
where $Y^{\mu\nu}$ and $Z^{\mu\nu}$ are antisymmetric tensors in spacetime while a one-form and a zero-form in solution space, respectively.\footnote{Note that the decomposition of $Y$ and $Z$ is not unique.} The $Z$ component represents the overlap between the $W$ and $Y$ freedoms.  The key point is that $Y$ and $Z$ do not influence the boundary conditions and, consequently, do not affect the action principle. Therefore, to ensure a well-defined action principle, we require the symplectic potential to vanish on causal boundaries, up to corner terms (see Subsection~\ref{sec:bdry-vs-corner} for further details).

We conclude this section with some comments on the holographic interpretation of the freedoms in the covariant phase space formalism. As discussed, the $W$-freedom can be interpreted as the generator of changes in the bulk boundary conditions. From the holographic perspective, this freedom in the saddle point approximation corresponds to multitrace deformations of the boundary theory \cite{Witten:2001ua} (see also \cite{Parvizi:2025shq} for a proof). The $Y$-freedom, on the other hand, does not appear in the bulk action and does not alter the bulk action principle. It can be interpreted as the symplectic potential of the boundary theory \cite{Parvizi:2025shq}. In other words, it specifies the sources and responses in the boundary theory and directly influences the boundary action principle.

\subsubsection{Boundary conditions, initial conditions and corner terms}\label{sec:bdry-vs-corner}
Boundary conditions are typically fixed  by requiring vanishing of the symplectic potential $\Theta$. As reviewed, $\Theta$ is specified by the choice of the boundary Lagrangian, ${W}$- freedom in our terminology \cite{Parvizi:2025shq, Parvizi:2025wsg}. Thus, one can generically get the desired boundary conditions by the choice of boundary Lagrangians/${W}$-freedom. However, as reviewed, the boundary Lagrangians are only defined up to corner terms, $Y$-freedom in our terminology \eqref{Y-Z-freedom}. 
Noting that the corner terms are typically defined on codimension-two (compact) spacelike surfaces (which are constant time slices), the freedom in the choice of the corner contribution may be viewed as a choice of the \textit{initial data}, and not boundary conditions. In other words, $Y$ terms do not affect the boundary conditions. This provides more relaxed possibilities for the boundary conditions. Explicitly, \textit{boundary conditions are set by vanishing of symplectic potential, up to $Y$-freedom}. 

A suitable way to formulate a well-defined action principle, without explicitly referring to boundary terms, is to examine the symplectic form
\begin{equation}
    \tilde{\Omega}[\Sigma_r] = \Omega[\Sigma_r] +  \int_{\partial\Sigma_{r}}\, \d{}x_{\mu\nu}\, \delta Y^{\mu\nu}\, .
\end{equation}
As is evident, the symplectic two-form is independent of the boundary Lagrangian. We also noted that all terms in the symplectic two-form that are given by codimension-two integral, the ``corner-terms,'' can be absorbed into a $Y$-term. This allows us to formulate a well-defined action principle with minimal requirements on the variations of the dynamical fields:
\begin{equation}\label{definition-bc}
 \inbox{   \Omega[\Sigma_r] = 0 + \text{corner terms}\, .}
\end{equation}
To clarify the above statement, let us consider the symplectic potential with a generic form
\begin{equation}
\Theta(\Sigma_r) = \int_{\Sigma_r}  X \ \delta P \, .
\end{equation}
If  $X$ is a constant in time, the above can be written as 
\begin{equation}
\Theta[\Sigma_r] = \int_{\Sigma_{tr}}  X \ \delta {\cal P} \, , \qquad {\cal P}=\int \d{}t P\, .
\end{equation}
Here, $\Sigma_{tr}$ denotes a codimension-two spacelike surface, i.e. a constant-time slice on $\Sigma_r$, with $X$ and ${\cal P}$ defined on $\Sigma_{tr}$. Note that $X$ and ${\cal P}$ may have arbitrary coordinate dependence on $\Sigma_{tr}$. Thus, the above expression represents a corner term: $\Sigma_{tr}$ can be interpreted either as the spatial boundary of a constant-time slice $\Sigma_t$, or as a constant-time slice of the timelike boundary $\Sigma_r$. Its vanishing, therefore, sets the initial data for the boundary theory. From a complementary viewpoint, the same expression can be regarded as a $Y$-term, with $Y = \d{}(X \, \delta {\cal P})$. The contribution of this term to the symplectic two-form arises through a codimension-two piece: 
\begin{equation}
\begin{split}
\Omega[\Sigma_r]
= \int_{\Sigma_{tr}} \delta X \wedge \delta \mathcal{P}\,. 
\end{split}
\end{equation} 
The final expression takes the form of a codimension-two integral, i.e., a corner term. In this sense, $X$ and $\mathcal{P}$ should be viewed as part of the initial data, and there is no reason to freeze them \cite{Harlow:2019yfa, Parvizi:2025shq, Parvizi:2025wsg}. In the following sections, we will further examine the role of the $Y$-term in relaxing boundary conditions, illustrated through various examples.

\subsubsection{Freelancing: arbitrary boundary conditions at arbitrary constant \texorpdfstring{$r$}{r} surfaces}\label{sec:bdry-bc-free-section}
We are now ready to apply the covariant phase space formalism (CPSF) to $3d$ Einstein gravity with arbitrary boundary conditions on $\Sigma_r$.  We take the boundary term as follows \cite{Parvizi:2025shq, Parvizi:2025wsg}
\begin{equation}
   W:= n_{\mu} W^{\mu} = - \frac{1}{2\kappa} \sqrt{-h}\, (w K + 2\mathrm{b} \ell^{-1}  )\, ,
\end{equation}
where $w$ and $\mathrm{b}$ are two constants. Including these boundary terms, the Dirichlet bulk action \eqref{action-AdS-3} is modified as follows: 
\begin{equation}\label{bulk-action-W-2}
S^{\text{\tiny{W}}}_{\text{\tiny{bulk}}}[\mathcal{M}_r] = \frac{1}{2\kappa} \int_{\mathcal{M}_r} \sqrt{-g} \left(\mathscr{R} + \frac{2}{\ell^2}\right) + \frac{1}{\kappa} \int_{\Sigma_r} \sqrt{-h}\,  \Big[\Big(1-\frac{w}{2}\Big)K - (1+\mathrm{b})\ell^{-1}\Big]\, .
\end{equation}
While in general $w$ and $b$ are independent constant parameters, in what follows we mainly focus on the two classes generically mentioned in the literature:
\begin{itemize}
\item {Unrenormalized boundary conditions,} that are
 specified by setting $\mathrm{b} = -1$ \cite{Liu:2024ymn}. In this case, the boundary term involves only the geometric quantity $K$. However, these boundary terms yield a divergent on-shell action in the asymptotic limit. 
\item{Renormalized boundary conditions,}
that are specified by $\mathrm{b} = -w$ (see section 6 of \cite{Parvizi:2025shq}, where $w$ was denoted as $k$) under which the action \eqref{bulk-action-W-2} takes the form
\begin{equation}\label{bulk-action-W-3}
\begin{split}
    S^{\text{\tiny{W}}}_{\text{\tiny{bulk}}}[\mathcal{M}_r] & =  S^{\text{\tiny{D}}}_{\text{\tiny{bulk}}}[\mathcal{M}_r] + \frac{w}{2}\int_{\Sigma_r} \sqrt{-h}\, \mathcal{T}\, .
\end{split}
\end{equation}
For $w = 0$, we recover the renormalized Dirichlet boundary condition. For $w = 1$, the renormalized Neumann and conformal boundary conditions emerge. 
\end{itemize}

A distinctive feature of $3d$ Einstein gravity is that both the Neumann and conformal boundary conditions arise from the same boundary term. In this case, the corresponding symplectic potentials take the following form:
\begin{equation}\label{NC-Theta}
    \begin{split}
        \Theta_{\text{\tiny{N/C}}}(\Sigma_r) & = \frac{1}{2} \int_{\Sigma_r} h_{ab}\, \delta(\sqrt{-h}\, \mathcal{T}^{ab})  \\
        & = -\frac{1}{2} \int_{\Sigma_r} \sqrt{-h}\left( \sqrt{-h}\, {\cal T}^{ab}\, \delta\hat{h}_{ab} -\delta\mathcal{T}\right)\, ,
    \end{split}
\end{equation}
where $\hat{h}_{ab}:=\frac{h_{ab}}{\sqrt{-h}}$ determines the boundary metric up to a conformal class. As \eqref{NC-Theta} shows, the N/C label here corresponds to Neumann and conformal boundary conditions, which are respectively defined by \cite{Parvizi:2025shq}\footnote{Note that in our definition of conformal boundary conditions we impose \(\delta \mathcal{T} = 0\)~\cite{Parvizi:2025shq, Parvizi:2025wsg, Taghiloo:2025oeu}, which we refer to as the {renormalized conformal boundary condition}. By contrast, the definition commonly used in the literature \cite{Anderson:2006lqb, anderson2008boundary, Witten:2018lgb, Coleman:2020jte, 
An:2021fcq, Anninos:2023epi, Anninos:2024wpy, Anninos:2024xhc, Liu:2024ymn, Banihashemi:2024yye} is \(\delta K = 0\), which we call the {unrenormalized conformal boundary condition}.} 
\begin{equation}\label{NC-bc-def}
    \begin{split}
        \text{Neumann bc:}& \qquad  \delta \big(\sqrt{-h}\, \mathcal{T}^{ab} \big)=0\, ,  \\
         \text{Conformal bc:}& \qquad \delta\left(\frac{h_{ab}}{\sqrt{-h}}\right)=0\, , \quad \& \quad \delta{\cal T}=0\, .
    \end{split}
\end{equation}
Unlike the unrenormalized class of boundary conditions, these ensure a finite on-shell action and finite physical quantities in the asymptotic limit. We therefore refer to them as renormalized boundary conditions. Unless otherwise stated, in this work will adopt renormalized boundary conditions.

We close this subsection with some further comments:
\begin{enumerate}
    \item 
    \textbf{General form of $W$.}  
    The boundary term \(W\) is, in general, a scalar functional \(W[h_{ab}, \mathcal{T}^{ab}]\). On a 2-dimensional boundary, the only independent algebraic scalars that can be constructed from \(h_{ab}\) and \(\mathcal{T}^{ab}\) are the trace \(\mathcal{T}\) and the T\(\bar{\text{T}}\) operator \(\Ottbar\). There also exist non-algebraic scalars involving derivatives of \(h_{ab}\) and \(\mathcal{T}^{ab}\). Recalling \eqref{EoM-ss} and \eqref{EoM-sa}, which are differential relations between \(h_{ab}\) and \(\mathcal{T}^{ab}\), all such scalars can ultimately be expressed as functions of \(\mathcal{T}\), \(\Ottbar\), or their derivatives. Restricting to purely algebraic combinations, \(W\) can be viewed as a functional of the form \(W = W(\mathcal{T}, \Ottbar)\) \cite{Ebert:2023tih}.

    \item 
    \textbf{Constraints on boundary conditions.}  
    Boundary conditions in gravity are subject to restrictions: they must be consistent with the Hamiltonian and momentum constraints. Consequently, the specification of boundary conditions must be accompanied by the enforcement of these constraints to ensure the overall consistency of the theory (see subsection \ref{sec: const-bc} for more details).

    \item 
    \textbf{Bulk boundary conditions as multitrace deformations.}  
    In the holographic framework, within the saddle-point approximation, modifying the bulk boundary conditions corresponds to deforming the boundary theory by multitrace operators \cite{Witten:2001ua} (see also \cite{Parvizi:2025shq} for a proof in the context of freelance holography). In this language, the role of the Hamiltonian and momentum constraints becomes transparent: they guarantee the covariance and (when appropriate) conformality of the boundary theory. Put differently, these constraints restrict admissible boundary multitrace deformations to those that preserve boundary covariance and conformality.\footnote{At the asymptotic AdS boundary, the boundary theory is conformal; at a finite radial cutoff, the trace of the stress tensor is non-vanishing, corresponding to a deformation by the double-trace T\(\bar{\text{T}}\) operator.}
\end{enumerate}

\subsubsection{Symplectic potential flow}\label{sec:symp-pot-flow}

Moving from asymptotic infinity of AdS space, besides the deformation of the dual boundary field theory, also induces an evolution on the boundary conditions on the bulk fields \cite{Parvizi:2025wsg, Adami:2025pqr}. To see the latter,  consider the on-shell variation of the bulk Lagrangian
\begin{equation}
    \delta \mathcal{L}^{\text{\tiny{W}}}_{\text{\tiny{bulk}}} = \partial_{\mu} \Theta^{\mu}_{\text{\tiny{W}}}\, ,
\end{equation}
where $\mathcal{L}^{\text{\tiny{W}}}_{\text{\tiny{bulk}}}$ is the Lagrangian density compatible with the $W$ boundary conditions. Let us now perform the integration over the codimension-one hypersurface $\Sigma_r$
\begin{equation}
    \frac{\d{}}{\d{}r} \Theta^{\text{\tiny{W}}}[\Sigma_r] = \delta \int_{\Sigma_r} \mathcal{L}^{\text{\tiny{W}}}_{\text{\tiny{bulk}}}\, , \qquad  \Theta^{\text{\tiny{W}}}[\Sigma_r] := \int_{\Sigma_r} n_{\mu} \Theta^{\mu}_{\text{\tiny{W}}}\, .
\end{equation}
This equation governs the radial evolution of the symplectic potential for arbitrary boundary conditions. Its global form is given by:
\begin{equation}
    \Theta^{\text{\tiny{W}}}[\Sigma_{r_2}] - \Theta^{\text{\tiny{W}}}[\Sigma_{r_1}] = \delta \int_{r_1}^{r_2} \d{}r \int_{\Sigma_r} \mathcal{L}^{\text{\tiny{W}}}_{\text{\tiny{bulk}}}\, .
\end{equation}
This result provides the first indication that boundary conditions evolve. In other words, if a $W$-type boundary condition is imposed at $\Sigma_{r_1}$, it will generally no longer retain the $W$-type form at $\Sigma_{r_2}$. The change in boundary conditions is driven by the bulk on-shell action.

We now compute the explicit form of the symplectic potential flow compatible with the renormalized boundary conditions \eqref{bulk-action-W-3}
\begin{equation}
    \mathcal{L}^{\text{\tiny{W}}}_{\text{\tiny{bulk}}} \Big|_{\text{\tiny{on-shell}}} = -\frac{2}{\kappa \ell r}\sqrt{-h} +\partial_{r} \left\{ \sqrt{-h} \left[\Big(-1+\frac{w}{2}\Big)\mathcal{T} + \frac{1}{\ell \kappa} \right] \right\} + \text{total boundary derivative}\, ,
\end{equation}
where total boundary derivative terms drop out when performing the integration over $\Sigma_r$.
To simplify this result, we use (see appendix \ref{appen:useful-rel} for more details),
\begin{equation}
    \partial_{r} \sqrt{-h} = \frac{2 \sqrt{-h}}{r} \left( 1- \frac{\ell \kappa}{2}\mathcal{T} \right) \, , \qquad \partial_{r}(\sqrt{-h}\, \mathcal{T}) = \frac{\ell \kappa}{r} \sqrt{-h}\, \Ottbar \, .
\end{equation}
Then, the bulk on-shell action becomes
\begin{equation}
    \begin{split}
        \mathcal{L}^{\text{\tiny{W}}}_{\text{\tiny{bulk}}} \Big|_{\text{\tiny{on-shell}}} & =  -\frac{\sqrt{-h}}{r} \left[ \mathcal{T} + \Big(1 - \frac{w}{2} \Big)\ell \kappa\, \Ottbar\right] \\
        & = \frac{\sqrt{-h}}{r} \left[ \frac{c}{24} \mathcal{R} - \frac{6\pi \ell^2}{c}(1-w) \Ottbar\right]\, ,
    \end{split}
\end{equation}
where in the last line we have used the Hamiltonian constraint \eqref{EoM-ss}. The flow of the symplectic potential, which contains the information about the evolution of boundary conditions as we move in $r$, is given by
\begin{equation}\label{symp-pot-RG}
 \inbox{  r \frac{\d{}}{\d{}r} \Theta^{\text{\tiny{W}}}[\Sigma_r] = - \frac{6\pi \ell^2}{c}(1-w)\, \delta \int_{\Sigma_r} \sqrt{-h}\,   \Ottbar\, ,}
\end{equation}
where we  used the fact that in $2d$ $\delta(\sqrt{-h}\,\mathcal{R})=0$ (up to corner terms). Some comments are in order:
\begin{enumerate}
    \item The flow of boundary conditions is sourced by the seminal T$\bar{\text{T}}$ operator.
\item  For $w=1$, the symplectic potential becomes non-flowing. Interestingly, $w=1$ corresponds to well-known boundary conditions: as noted below \eqref{bulk-action-W-3}, it describes the renormalized Neumann and conformal boundary conditions. Consequently, the distinctive feature of the $3d$ Neumann and conformal symplectic potentials is
\begin{equation}\label{non-flow-N-C}
    \Theta_{\text{\tiny{N/C}}}(\Sigma_r) = \Theta_{\text{\tiny{N/C}}}(\Sigma)\, .
\end{equation}
Explicitly, for Neumann (N) and Conformal (C) symplectic potentials, we have 
\begin{equation}\label{N-Conf-No-running}
   \begin{split}
 &\text{N. symp. pot. :}  \quad \int_{\Sigma_r} \delta(\sqrt{-h}\, \mathcal{T}^{ab})\ h_{ab} =  \int_{\Sigma} \delta(\sqrt{-q}\, {T}^{ab} )\ q_{ab}\, , \\
& \text{C. symp. pot. :} \quad \int_{\Sigma_r} \sqrt{-h}\left( \delta\mathcal{T}-\sqrt{-h}{\cal T}^{ab} \delta\hat{h}_{ab}\right) =  \int_{\Sigma} \sqrt{-q}\left(\delta{T}- \sqrt{-q}{T}^{ab} \delta\hat{q}_{ab}\right)
   \end{split}
\end{equation}
where in the second line 
\begin{equation}\label{det-free-def}
\hat{q}_{ab}:=\frac{q_{ab}}{\sqrt{-q}}\, , \qquad \hat{h}_{ab}:=\frac{h_{ab}}{\sqrt{-h}}\, .    
\end{equation}
This result demonstrates that the Neumann and conformal symplectic potentials remain unchanged under radial evolution; in other words, they are RG invariant. We again note that this result is only valid in $3d$ gravity and not in higher-dimensional cases. 
\item While conformal or Neumann boundary conditions at a constant-$r$ surface imply $\Theta_{\text{\tiny{N/C}}}(\Sigma_{r})=0$, this condition is not unique to those choices: $\Theta_{\text{\tiny{N/C}}}(\Sigma_{r})=0$ admits other solutions beyond the standard Neumann or conformal boundary conditions. Thus, in principle, one can move between different families of boundary conditions satisfying $\Theta_{\text{\tiny{N/C}}}=0$ as the radial coordinate $r$ is varied. We will illustrate this point with explicit examples in the following sections.
\end{enumerate}

\subsection{Integrating effective action of boundary theory at arbitrary constant \texorpdfstring{$r$}{r} surfaces}\label{sec:bdry-bc-free-solnspace}
Given the above setup, one can derive the flow equation for the bulk on-shell (or equivalently boundary) action and integrate it to evaluate its value at an arbitrary radial slice $r$ \cite{Parvizi:2025wsg}. The resulting on-shell action depends on the choice of boundary conditions. To begin, let us consider Dirichlet boundary conditions. Under an arbitrary diffeomorphism generated by $\xi$, the variation of the Einstein–Hilbert action with Dirichlet boundary conditions takes the form
\begin{equation}
    \delta_{\xi} S^{\text{\tiny{D}}}_{\text{bulk}}[\mathcal{M}_r] \Big|_{\text{on-shell}} = -\frac{1}{2} \int_{\Sigma_r} \sqrt{-h}\, \mathcal{T}^{ab}\, \delta_{\xi} h_{ab}\, . 
\end{equation}
To achieve the radial evolution of the bulk on-shell action, we choose $\xi = \partial_{r}$, then 
\begin{equation}\label{radial-bulk-action}
   \begin{split}
       r\frac{\d{}}{\d{}r} S^{\text{\tiny{D}}}_{\text{bulk}}[\mathcal{M}_r] \Big|_{\text{on-shell}} & = -\frac{1}{2} \int_{\Sigma_r} \sqrt{-h}\, {\mathcal{T}}^{ab}\, r\partial_{r} h_{ab} \\
       & = -   \int_{\Sigma_r} \sqrt{-h} \left( \mathcal{T}+ \frac{12\pi \ell^2}{c} \Ottbar \right)\, \\
       &= r \frac{\d{}}{\d{}r} {S}^{\text{\tiny{D}}}_{\text{bdry}}[\Sigma_r]\, . 
   \end{split}
\end{equation}
In the second line of the above, we used \eqref{hab-r-T} and \eqref{ottbar}, and in the last line used the expression for finite-distance holography proposal in the saddle point approximation, as established in \cite{Parvizi:2025wsg}
\begin{equation}\label{Finite-cutoff-Holography-D}
    S^{\text{\tiny{D}}}_{\text{bulk}}[\mathcal{M}_r] \Big|_{\text{on-shell}} = {S}^{\text{\tiny{D}}}_{\text{bdry}}[\Sigma_r]\, ,
\end{equation}
where the RHS denotes the boundary action living on $\Sigma_r$ dual to the bulk theory $\mathcal{M}_r$. Eq. \eqref{radial-bulk-action} is a first-order differential equation in $r$ for ${S}^{\text{\tiny{D}}}_{\text{bdry}}[\Sigma_r]$ and defines a deformation flow equation for the boundary theory. This first-order differential equation can be integrated, once we specify a boundary or initial condition, say at $r\to\infty$; for example, 
\begin{equation}
    \lim_{r\to \infty}{S}^{\text{\tiny{D}}}_{\text{bdry}}[\Sigma_r] =  {S}^{\text{\tiny{D}}}_{\text{bdry}}[\Sigma]\, . 
\end{equation}
Here, the right-hand side represents the boundary action evaluated at the asymptotic AdS boundary $\Sigma$, where it is assumed to be defined via the standard gauge/gravity correspondence.

Eq.~\eqref{radial-bulk-action} depends on $\mathcal{T}$ and $\Ottbar$, which are related via \eqref{All-r-quantities-2} to their asymptotic counterparts $\text{T}$ and $\text{T}\overline{\text{T}}$ at $r \to \infty$. Employing \eqref{EoM-ss}, the deformation flow equation \eqref{radial-bulk-action} can be recast as 
    \begin{equation}  
      \inbox{r\frac{\d{}}{\d{}r} {S}^{\text{\tiny{D}}}_{\text{bdry}}[\Sigma_r] =  {\frac{c }{24\pi}}\int_{\Sigma_r} \text{d}^2x\sqrt{-h} \left[  {\cal R} - \big(\frac{12\pi\ell }{c}\big)^2 {\Ottbar}  \right]\, .}
    \end{equation} 
Applying the Gauss–Bonnet theorem, we can rewrite the equation as follows:
   \begin{equation}\label{2d-TTbar}       r \frac{\d{}}{\d{}r} {S}^{\text{\tiny{D}}}_{\text{bdry}}[\Sigma_r] = \frac{c}{12}  \chi_{_{\text{\tiny{E}}}} -\frac{6\pi\ell^2 }{c}\int_{\Sigma_r} \text{d}^2x\ \sqrt{-h}\,  {\Ottbar}\, , 
    \end{equation} 
where $\chi_{_{\text{\tiny{E}}}}$ denotes the Euler characteristic of the boundary surface $\Sigma_r$. Using \eqref{r-dep-ttbar-3}, we can integrate the flow  \eqref{2d-TTbar}, yielding:
\begin{equation}\label{Integrated-effective-action-D} 
      \inbox{ {S}^{\text{\tiny{D}}}_{\text{bdry}}[\Sigma_r] =  {S}^{\text{\tiny{D}}}_{\text{bdry}}[\Sigma] + \frac{c \chi_{_{\text{\tiny{E}}}}}{12} \ln \left( \frac{r}{r_\infty} \right) + \frac{3\pi\ell^4}{c r^2} \int \text{d}^2x \sqrt{-q}\, \ttbarb\, .  }
\end{equation}
The above gives the expression for the renormalized on-shell effective action of the $2d$ CFT in terms of the cutoff scale $\Lambda:=r^2/\ell^3$. The $\log$-term is the expected expression coming from the trace anomaly (an integrated version of that). The logarithmic term, and that $\ln\left( \frac{r}{r_\infty} \right)$ is a monotonically decreasing function of $r$, is a manifestation of the standard $c$ theorem \cite{Zamolodchikov:1986gt}.

The $1/r^2$ term, which falls as $1/\Lambda$, is the seminal statement that the T$\bar{\text{T}}$ deformation of the boundary CFT in $\text{AdS}_3$ admits a dual interpretation as finite cutoff holography \cite{Guica:2019nzm}.  If $\ttbarb\geq 0$, this term is monotonically decreasing. 
We now derive this result within the framework of freelance holography \cite{Parvizi:2025wsg}. It is worth mentioning that, while our derivation is carried out in the saddle-point approximation (corresponding to the large-$N$ limit), the duality proposed in \cite{Guica:2019nzm} is conjectured to hold even at finite $N$. This proposal is supported by Zamolodchikov’s factorization formula for the $2d$ T$\bar{\text{T}}$ deformation at finite $N$ \cite{Zamolodchikov:2004ce} (see also \cite{Jiang:2019epa} for a review).

We close by the comment that \eqref{Integrated-effective-action-D} has been written in terms of asymptotic boundary data. To interpret it as a boundary theory on $\Sigma_r$, we rewrite it in terms of intrinsic data on $\Sigma_r$ via \eqref{r-dep-ttbar-3}
\begin{equation}\label{D-action-integrated} 
      \inbox{ {S}^{\text{\tiny{D}}}_{\text{bdry}}[\Sigma_r] =  {S}^{\text{\tiny{D}}}_{\text{bdry}}[\Sigma] + \frac{c\, \chi_{_{\text{\tiny{E}}}} }{12} \ln \left( \frac{r}{r_\infty} \right) + \frac{3\pi\ell^2}{c} \int_{\Sigma_r} \text{d}^2x \sqrt{-h}\, \Ottbar\, .  }
\end{equation}

\section{A systematic analysis of consistent boundary conditions}\label{sec:classification-bc}
So far, we have discussed three types of boundary conditions commonly appearing in the literature: Dirichlet, Neumann, and conformal. In this section, we develop a formal framework to introduce and systematically study additional kinds of ``renormalized boundary conditions'' at arbitrary $r$.

To proceed, we note that  the symplectic two-form is $r$ independent, up to codimension-two ``boundary terms'', i.e. 
\begin{equation}\begin{split}
\Omega &:= -\frac{1}{2} \int_{\Sigma_r}  \delta (\sqrt{-h}\, \mathcal{T}^{ab}) \wedge \delta h_{ab} + \text{corner terms} \\ &= -\frac{1}{2} \int_{\Sigma}  \delta (\sqrt{-q}\, {T}^{ab}) \wedge \delta q_{ab}+ \text{corner terms} \,  .
\end{split}
\end{equation}
The above equality follows directly from \eqref{EoM-GR-mat-decompose}.
A necessary requirement for the gauge/gravity correspondence (and more generally, the AdS/CFT duality) at finite cutoff is that there should be no symplectic flux through the timelike boundary of the cutoff AdS space. Moreover, consistency of the correspondence also requires the vanishing of the codimension-one part of the symplectic form. This latter is fulfilled by the choice of boundary conditions, explicitly \eqref{definition-bc},  
\begin{equation}\label{def-bc}
\Omega = 0 + \text{corner terms}\, .
\end{equation}
Conversely, any boundary condition that satisfies the above condition is a good, i.e. classically consistent, boundary condition for the correspondence/duality. The above also shows that if a boundary condition is available at a given radius $r_1$, it is also consistent at any other radius $r_2$, provided that, in general, one should deform the dual field theory by the appropriate double-trace deformation to make up for the evolution in $r$. For the standard T$\bar{\text{T}}$-deformation literature and the case of $3d$ pure gravity, one is prescribed to fulfill \eqref{def-bc}  by the choice of Dirichlet boundary condition at any given $r$, which amounts to deforming the $2d$ CFT side with T$\bar{\text{T}}$ operator. However, as we discussed, one can fulfill \eqref{def-bc} by a variety of other boundary conditions, guaranteed by the addition of appropriate choices of boundary Lagrangians (besides GHY)  to \eqref{action-AdS-3} \cite{Parvizi:2025shq, Parvizi:2025wsg}, see \eqref{D-action-integrated} for an explicit derivation. However, as we have established for all of these cases one should still deform the CFT side by T$\bar{\text{T}}$ operator, but with a different numerical coefficient, cf. \eqref{symp-pot-RG}.

\subsection{Covariant boundary conditions}

The Dirichlet, Neumann, and conformal boundary conditions clearly fulfill \eqref{def-bc}; however, these boundary conditions are by no means exhaustive of all the possibilities. For example, recalling that 
\begin{equation}
 h^{ab}\delta h_{ab}=\delta(\ln\det{h}) \quad \Longrightarrow \quad \delta \big(\mathcal{F}({-h})\, h^{ab}\big)\wedge \delta h_{ab}=0\, ,  
\end{equation}
for any arbitrary function of $\mathcal{F}$, one general class of boundary conditions can be given by 
\begin{equation}
    \delta(\sqrt{-h}\, {\cal T}^{ab})= X \delta(\mathcal{F}({-h})  h^{ab})\, ,
\end{equation}
for an arbitrary constant $X$ ($\delta X=0$). This family of boundary conditions interpolate between Neumann for $X=0$ and Dirichlet for $X\gg 1$. 

As another example we note that \cite{Parvizi:2025shq} 
\begin{equation}
    \delta (\sqrt{-h}\, \mathcal{T}^{ab}) \wedge \delta h_{ab}= \delta (\sqrt{-h}{\cal T})\wedge \delta\ln\sqrt{-h} + \delta (\sqrt{-h}\,\hat{h}_{ab})\wedge\delta\big(\sqrt{-h}\hat{{\cal T}}^{ab}\big)\, , 
\end{equation}
where $\hat{h}_{ab}$ is defined in \eqref{det-free-def} and
\begin{equation}
    \hat{{\cal T}}^{ab}:= {\cal T}^{ab}-\frac12 {\cal T} h^{ab}\, ,
\end{equation}
is the traceless part of rBY-EMT. Recalling that 
\begin{equation}
    \hat{h}^{ab}\delta {\hat h}_{ab}=0 \quad \Longrightarrow \quad \delta({\cal F}(-h)\ \hat{h}^{ab})\wedge \delta {\hat h}_{ab}=0\, , 
\end{equation}
for an arbitrary function $\mathcal{F}$, the above yields the boundary conditions \footnote{Note that while $\hat{\cal T}^{ab}$ is traceless, $\delta \hat{\cal T}^{ab}$ need not be traceless.}
\begin{equation}\label{Interpolating-class}
    \delta(\sqrt{-h} {\cal T})= X \delta{\sqrt{-h}}\, ,\qquad  \delta ({\cal F}{(-h)} \hat{h}^{ab})=Y \delta\big(\sqrt{-h}\,\hat{{\cal T}}^{ab}\big)\, ,
\end{equation}
where $X=X(\sqrt{-h}, {\cal T})$ and $Y$ is an arbitrary constant. 
These boundary conditions interpolate between Dirichlet boundary conditions ($X=0, Y=0$), Neumann boundary conditions ($X=0, Y\gg 1$) and conformal boundary conditions ($Y=0, X={\cal T}$). 
\subsection{Non-covariant boundary conditions}\label{sec:non-covariant-bc}
In pure gravity, the relevant boundary data involve two covariant tensors, ${\cal T}^{ab}$ and $h_{ab}$, as well as tensor densities such as $\sqrt{-h}$, and a scalar ${\cal T}$. Up to tensor densities, we have so far considered boundary conditions that preserve the $2d$ covariance in a systematic way, working with the combinations $\sqrt{-h}\,\hat{{\cal T}}^{ab}$, ${\cal T}$, $\sqrt{-h}$, and $\hat{h}_{ab}$ \cite{Parvizi:2025shq, Parvizi:2025wsg}. To explore additional classes of boundary conditions, however, it becomes necessary to split the boundary indices, which explicitly breaks boundary covariance \footnote{{Non-covariant boundary conditions arise in various contexts. One such situation occurs when we consider a gravity theory coupled to matter fields with standard covariant boundary conditions. Upon integrating out the matter fields, the boundary conditions for the gravitational sector may change, typically yielding non-covariant boundary conditions \cite{Draper:2025kcr}.}}. 

To carry this out systematically, we introduce two fixed non-dynamical vector fields, $t^a$ and $k^a$
\begin{equation}
    \delta t^{a} = 0\, , \qquad \delta k^{a} = 0\, .
\end{equation}
The vanishing variation of $t^{a}$ indicates that it defines a preferred background direction.\footnote{A non-varying vector field parallels the notion of a Killing vector field (though on the solution space). In the AdS$_3$ gravity, one may show that all solutions at least have two globally defined Killing vectors \cite{Sheikh-Jabbari:2014nya}.} Using this vector field, we can impose additional non-covariant boundary conditions. 
In terms of these two independent background vector fields, one can always write
\begin{equation}
\begin{split}
h_{ab} &= x\, t_{a}t_{b} + y\, (t_{a} k_{b} + k_{a} t_{b}) + z\, k_{a}k_{b}\, ,   \\
\mathcal{T}^{ab} &= X\, t^{a} t^{b} + Y\, (t^{a}k^{b} + t^{b} k^{a}) + Z\, k^{a} k^{b}\, .
\end{split}
\end{equation}
It is worth emphasizing that the construction is valid for any choice of vector fields: the two background vectors $t^{a}$ and $k^{a}$ may be timelike–spacelike, null–null, timelike–null, or spacelike–null.

The complete information of $h_{ab}$ and $\mathcal{T}^{ab}$ is encoded in the six scalars $(x, y, z)$ and $(X, Y, Z)$, or equivalently in the six quantities $(t \cdot t,\, t \cdot k,\, k \cdot k)$ and $(\mathcal{T}_{tt},\, \mathcal{T}_{tk},\, \mathcal{T}_{kk})$, with
\begin{equation}
    \mathcal{T}_{tt} = t^{a} t^{b}  \mathcal{T}_{ab}\, , \qquad \mathcal{T}_{tk} = t^{a}k^{b} \mathcal{T}_{ab}\, , \qquad \mathcal{T}_{kk} = k^{a} k^{b} \mathcal{T}_{ab}\, .
\end{equation}
These two sets are connected through
\begin{equation}
    x = \frac{k \cdot k}{h}
    \, , \qquad 
    y= -\frac{t \cdot k}{h} 
    \, , \qquad z =
    \frac{t \cdot t}{h} 
    \, ,
\end{equation}
and 
\begin{equation}\label{X-Y-Z}
   \begin{split}
       & X = \frac{(t \cdot k)^2 \mathcal{T}_{kk} - 2 (k\cdot k)(t \cdot k) \mathcal{T}_{tk}+(k\cdot k)^2 \mathcal{T}_{tt}}{(-h)^2}
       \, , \\
       & Y = \frac{- (t\cdot t)(t \cdot k) \mathcal{T}_{kk} + (k \cdot k)(t \cdot t)\mathcal{T}_{tk} + (t \cdot k)^2 \mathcal{T}_{tk}-(k\cdot k)(t \cdot k)\mathcal{T}_{tt}}{(-h)^2}
       \, ,  \\
       & Z = \frac{(t \cdot t)^2 \mathcal{T}_{kk} - 2 (t\cdot t)(t \cdot k) \mathcal{T}_{tk}+(t\cdot k)^2 \mathcal{T}_{tt}}{(-h)^2}
       \, ,
   \end{split}
\end{equation}
where \footnote{Here we have fixed $\epsilon_{ij} t^{i} k^{j}=1$.}
\begin{equation}\label{det-h-t-k}
    h:=\det{h_{ab}} = (k\cdot k)(t \cdot t) -(t\cdot k)^2 \, .
\end{equation}
and 
\begin{equation}
    {\cal T} = X (t \cdot t) + 2Y (t\cdot k) + Z (k \cdot k), \qquad \Ottbar=(-2h) (XZ-Y^2)\, .
\end{equation}
Finally, the symplectic form takes the form
\begin{equation}\label{non-covariant-bc}
\begin{split}
        \Omega[\Sigma] & =\int_{\Sigma_r} \delta \left(\sqrt{-h}\, \mathcal{T}^{ab} \right)\,\wedge  \delta h_{ab}\\
        & = \int_{\Sigma_r}\left[ \delta (\sqrt{-h}\, X) \wedge \delta(t \cdot t) +2  \delta(\sqrt{-h}\, Y) \wedge \delta(t \cdot k) +\delta(\sqrt{-h}\, Z) \wedge \delta (k \cdot k) \right]\, .
\end{split}
\end{equation}
This expression gives rise to three canonical pairs. 

By fixing (i.e., setting to zero) one variable from each pair, we obtain eight possible boundary conditions, three of which correspond explicitly to the standard Dirichlet, Neumann and conformal cases 
\begin{subequations}
\begin{align}
   \text{Dirichlet boundary condition:}&\qquad  \delta(t \cdot t)=\delta(t \cdot k)=\delta(k \cdot k)=0\, ,\\
   \text{Neumann boundary condition:}&\qquad  \delta (\sqrt{-h}\, X)=\delta (\sqrt{-h}\, Y)=\delta (\sqrt{-h}\, Z)=0\, ,\\ 
   \text{Conformal boundary condition:}&\qquad  t\cdot t:=0,\ k\cdot k:=0,\ \delta(\sqrt{-h}\, Y)=0\, .\label{conf-bc-sec32}
\end{align}    
\end{subequations}
In the last line we used \eqref{det-h-t-k}, $t\cdot t:=0$ means $t\cdot t=0, \delta(t\cdot t)=0$, and similarly for $k$ and that for this case ${\cal T}=2(t\cdot k)Y=2\sqrt{-h} Y$. Note also that for \eqref{conf-bc-sec32} $h_{ab}= (t_a k_b+ t_b k_a)/(t\cdot k)$. Out of the remaining five cases, considering the $t, k$ exchange symmetry, there are only two independent cases. We discuss three examples of this family  in section \ref{sec:noncovariant-bc-solnspace}. 

In Sections \ref{sec: bdry-constraints} and \ref{sec:noncovariant-bc-solnspace}, we will examine several illustrative examples. More general classes of boundary conditions can also be obtained through an analysis similar to \eqref{Interpolating-class}, but we will not pursue these extensions here.

\subsection{Constraints on boundary conditions}\label{sec: const-bc}
The boundary value problem in gravity is inherently subtle due to diffeomorphism invariance, which gives rise to the Hamiltonian and momentum constraints. Consequently, as noted in the second comment below \eqref{NC-bc-def}, any admissible boundary conditions must be formulated in a way that is consistent with these constraints. In this subsection, we briefly examine  these constraints in the context of AdS$_3$ gravity and how they affect the choice of boundary conditions, both at finite distance and at infinity.

\paragraph{Boundary conditions in a finite distance.}
Consider the Hamiltonian constraint at a finite distance \eqref{EoM-ss}. Multiplying it by $\sqrt{-h}$, integrating over $\Sigma_r$,  we obtain 
\begin{equation}
   \delta \int_{\Sigma_{r}} \sqrt{-h} \left(\mathcal{T} + {\frac{{c} }{24\pi}} {\cal R} + \frac{6\pi\ell^2}{{c}}\Ottbar \right)=0\, .
\end{equation}
$\int_{\Sigma_{r}} \sqrt{-h}\, {\cal R}$ is proportional to the Euler characteristic of $\Sigma_r$ which has vanishing variation. Therefore, 
\begin{equation}
   \delta \int_{\Sigma_{r}} \sqrt{-h} \left(\mathcal{T} + \frac{6\pi\ell^2}{{c}}\Ottbar \right)=0\, .
\end{equation}
This shows that, as a consequence of the Hamiltonian constraint, the trace and T$\bar{\text{T}}$ deformations are not independent, and any admissible boundary conditions must respect this relation.

\paragraph{Boundary condition at infinity.}
We now examine how the Hamiltonian constraint restricts the asymptotic boundary conditions. Starting from the trace anomaly equation \eqref{cons-trace}, we multiply by $\sqrt{-q}$, integrate over $\Sigma$, and then take the variation, yielding
\begin{equation}
    \delta \int_{\Sigma} \sqrt{-q} \left(T + \frac{c}{24 \pi} {R} \right) =0\, .
\end{equation}
As in the previous case, we have $\delta \int_{\Sigma} \sqrt{-q}\, R = 0$, and we are left with
\begin{equation}
    \delta \int_{\Sigma} \sqrt{-q}\, T =0\, .
\end{equation}
Therefore, all admissible asymptotic boundary conditions must be chosen so as to preserve the above condition. In what follows, we frequently  use this constraint.

\section{Solution space for various boundary conditions}\label{sec: bdry-constraints}
In subsection \ref{sec:integrability} (see also appendix \ref{appen:useful-rel} for more useful identities), we solved the radial evolution of the Einstein equations and determined the solutions in terms of the boundary data $\{q_{ab}, T_{ab}\}$, subject to the constraints in \eqref{cons-trace}. To obtain explicit solutions, one must further solve these constraint equations as well as imposing the desired boundary conditions. As we will see, different choices of boundary conditions give rise to distinct classes of solutions, each specified by two arbitrary functions of a single variable—that is, two functions defined on codimension-two surfaces.

To clarify this, let us perform a simple counting: the boundary data consist of six components, $\{q_{ab}, T^{ab}\}$. Imposing three boundary conditions—necessary for a well-posed variational principle—leaves us with three free components. Among the constraints, the Hamiltonian constraint is algebraic and fixes one of these remaining components. The momentum constraints, which are two differential equations, fix two more components. However, since these are first-order differential constraints, they leave residual freedom in the form of two arbitrary functions defined on codimension-two surfaces, two functions of one variable, i.e., corner data. Therefore,  regardless of the specific choice of boundary conditions, the remaining freedom in the solution space is generically encoded in \textit{two} arbitrary functions on \textit{codimension-two} surfaces.\footnote{One can always restrict oneself to a subspace of the solutions involving only one function of a single variable.}

In this section and the next, we restrict our study to solutions with flat $2d$ boundary, with various boundary conditions and the non-flat boundary cases are briefly discussed in appendix \ref{appen:non-flat-bc}.


\subsection{Dirichlet boundary conditions, Ba\~nados geometries}\label{sec:Banados}

Our first example involves Dirichlet boundary conditions imposed at the boundary $\Sigma$, which we take it to be flat. In this case, we fix the boundary metric by setting $\delta q_{ab} = 0$, i.e. 
\begin{equation}\label{flat-AdS-bdry}
       \d{}s_{q}^{2} =  q_{ab} \d{}x^{a} \d{}x^{b} = - \ell^2 \d{}x^{+} \d{}x^{-}\, .
\end{equation}
Since the boundary is flat, the trace anomaly \eqref{cons-trace} implies that the boundary energy-momentum tensor is traceless, $T = 0$. Moreover, the conservation equation $D_a T^{ab} = 0$ yields
\begin{equation}\label{flat-bdry-EMT}
        T_{\pm \pm} = - \frac{c}{12 \pi} L_{\pm} (x^{\pm})\, , \qquad T_{+-}=0\, ,
\end{equation}
where $L_{\pm} (x^{\pm})$ are two arbitrary functions of their arguments. If we take the boundary to be a cylinder, then we need to impose some periodicity condition on $L_{\pm} (x^{\pm})$, e.g., $L_{\pm} (x^{\pm})=L_{\pm} (x^{\pm}+2\pi)$. The holographic T$\bar{\text{T}}$ operator is  given by
\begin{equation}
    \ttbarb = \frac{c^2}{{18} \pi^2 \ell^4} L_+ L_-\, .
\end{equation}
With the boundary data in hand, the full on-shell $3d$ line element can be written as
\begin{equation}\label{Banados-3D}
        \d{}s^2 = \ell^2 \frac{\d{}r^2}{r^2} - \left( r \d{}x^+ - \ell^2 \frac{L_-}{r} \d{}x^-\right)\left( r \d{}x^- - \ell^2 \frac{L_+}{r} \d{}x^+\right)\, ,
\end{equation}
with
\begin{equation}
  \sqrt{-h} = \frac{1}{2r^2} \left( r^4- {\ell^4} L_{+}L_{-} \right)\, .
\end{equation}
The above are the class of Ba\~nados geometries \cite{Banados:1998gg}, see 
\cite{Sheikh-Jabbari:2014nya, Sheikh-Jabbari:2016unm, Sheikh-Jabbari:2016npa} for further analysis. These geometries are constructed so that they satisfy Dirichlet boundary conditions at the asymptotic AdS boundary $\Sigma$, namely, $\delta q_{ab}=0$.   

The conjugate EMT at an arbitrary $r$ for the Ba\~nados geometries is
\begin{equation}\label{Banados-calT}
   \sqrt{-h}\, \mathcal{T}^{\pm \pm} =  - \frac{c}{6 \pi} \frac{r^2 L_\mp}{r^4 - \ell^4 L_+ L_-}\, , \qquad \sqrt{-h}\, \mathcal{T}^{+-} = - \frac{c}{6 \pi}\frac{\ell^2 L_+ L_-}{r^4 -  \ell^4 L_+ L_-}\, ,
\end{equation}
with
\begin{equation}\label{calT-Babandos}
    \mathcal{T} = - \frac{c \ell^2}{{3} \pi} \frac{ L_+ L_- }{ r^4 - {\ell^4} L_{+}L_{-} }, \quad \Ottbar = \frac{c^2 }{{18} \pi^2} \frac{L_+ L_-}{{r^4} - \ell^4 L_{+}L_{-}}.
\end{equation}

\paragraph{Boundary condition flow.} Using equations in section \ref{sec:integrability} and appendix \ref{appen:useful-rel}, one can verify that,
\begin{equation}
    \sqrt{-q} q^{ab} = \sqrt{-h} \left( h^{ab} - \frac{12 \pi \ell^2}{c} \mathcal{T}^{ab} \right)\, .
\end{equation}
This equation yields,
\begin{equation}\label{D-into-DN}
    \delta(\sqrt{-h}\, {\cal T}^{ab})= -\frac{c\sqrt{-h}}{24\pi \ell^2 } {\cal G}^{abcd} \delta h_{cd}\, , \qquad {\cal G}^{abcd}:= h^{ac} h^{bd}+h^{ad} h^{bc}-h^{ab} h^{cd}\, ,
\end{equation}
where ${\cal G}^{abcd}$ is the Wheeler-deWitt metric \cite{PhysRev.160.1113}. The above shows how the Dirichlet bc at large $r$ evolves as we move to generic $r$, where we get a mixture of Dirichlet and Neumann bc's.\footnote{As a check for the analysis, we note that the left-hand-side of \eqref{D-into-DN} at large $r$ scales as $r^{-2}$ (or $r^{-4}$), cf. \eqref{Banados-calT}, while the right-hand-side goes as $r^0$. Therefore, at large $r$, the left-hand side should vanish, recovering the desired Dirichlet boundary conditions. Moreover, we note that \eqref{D-into-DN} may also be written as $ \delta(\sqrt{-h}\, {\cal T}^{ab})= -\frac{c\sqrt{-h}}{12\pi \ell^2} \delta (\sqrt{-h} h^{ab})$, in which the right-hand-side vanishes for a conformal boundary condition at radius $r$. \label{footnote-9}}

\paragraph{Surface charges and their algebra.} Instead of discussing the details of the geometry, one could have studied the algebra of charges associated with diffeomorphisms that move us within the given set of solutions, in this case, within the Ba\~nados class \eqref{Banados-3D}. This exercise has been worked through in \cite{Compere:2015knw}, where it was shown that the charge algebra consists of two copies of Virasoro algebra at Brown-Henneaux central charge \cite{Brown:1986nw}. As noted in \cite{Compere:2015knw}, the charges and the associated symmetry algebras are ``symplectic'' symmetries which can be recovered and computed at any arbitrary radius; they are not just ``asymptotic'' symmetries that can be computed at asymptotic regions of AdS$_3$. In a different wording, the class of Ba\~nados solutions form a solution phase space and one can move from one solution in the class to another  by the action of symplecto-morphisms which are coadjoint to subalgebras of $3d$ diffeomorphisms and form two copies of Virasoro algebra.

\subsection{Neumann \texorpdfstring{AdS$_3$}{AdS3} geometries} \label{sec:N-bc-infty}

\noindent 
As the next example we explore Neumann boundary conditions at $\Sigma$,  requiring $\delta (\sqrt{-q} \, T^{ab}) = 0$, with $T^{ab}$ taken to be covariantly constant. Without loss of generality, we may choose
\begin{equation}\label{N-bc-Sigma}
    \sqrt{-q}\, T^{ab} = \frac{c}{12 \pi \ell^2} \begin{bmatrix}
1 & 0 \\
0 & 0
\end{bmatrix}\, ,
\end{equation}
in $(t,\phi)$ basis. The metric $q_{ab}$, which we choose to be flat, but otherwise arbitrary, in this basis, takes the form, 
\begin{equation}
    \d{}s_{q}^2= q_{ab}  \d{}x^{a} \d{}x^b = 2 f \d{}t \d{}\phi + {t g} \d{}\phi^2\, ,
 \end{equation}
where $f=f(\phi)$ and $g=g(\phi)$ are periodic functions, consistent with the cylindrical topology of the boundary. The corresponding bulk line element is then given by
\begin{equation}\label{metric-N-bc-Sigma}
    \d{}s^{2} =  \ell^2\frac{\d{}r^2}{r^2} + \frac{ 2r^2}{\ell^2} f \d{}t \d{}\phi +\left( \frac{r^2}{\ell^2} t g -f \right) \d{}\phi^2\, ,
\end{equation}
for which $\sqrt{-h} = \frac{r^2}{\ell^2} f(\phi)$. The boundary EMT \eqref{N-bc-Sigma} has the following special features
\begin{equation}\label{N-bc-features-Sigma}
    T=0\, , \qquad \ttbarb =0\, , \qquad T_{ac}T^{cb}=0\, . 
\end{equation}
Using equations in section \ref{sec:integrability}, we find the rBY-EMT at a finite distance
\begin{equation}\label{EMT-N-r}
    \sqrt{-h}\, \mathcal{T}^{ab} = \frac{c}{12 \pi r^2} \begin{bmatrix}
1 & 0 \\
0 & 0
\end{bmatrix}\, .
\end{equation}
From this, we deduce $\delta(\sqrt{-h}\, \mathcal{T}^{ab})=0$, which confirms our general result \eqref{non-flow-N-C} that the Neumann boundary condition does not flow as we move in $r$. In addition, the special features of the boundary EMT \eqref{N-bc-features-Sigma} are also preserved under RG
\begin{equation}
    \mathcal{T}=0\, , \qquad \Ottbar =0\, , \qquad \mathcal{T}_{ac} \mathcal{T}^{cb}=0\, .
\end{equation}

\paragraph{Surface charges and their algebra.} In Appendix \ref{appen:Neumann-charges} we have shown that the symplectic symmetries which moves us within the solution space with Neumann boundary conditions is the $u(1)$ Kac-Moody algebra \eqref{charge-algebra-Neumann}.

As a final comment, the solution \eqref{metric-N-bc-Sigma} is a new solution not discussed in the literature before. A similar, but still different, solution in  a different (non-Feferman-Graham) gauge, has been previously obtained and discussed in \cite{Afshar:2016wfy, Afshar:2016kjj, Afshar:2016uax, Afshar:2017okz}. The associated boundary condition is particularly well-suited for describing soft hair on the near-horizon geometry of BTZ black holes as well as the fluid/gravity correspondence \cite{In-progress-freelance-hydro}. In contrast, Ba\~nados geometries are more appropriate for capturing the asymptotic behavior of soft (boundary) gravitons on BTZ backgrounds. In principle, one can construct interpolating geometries that bridge these two regimes \cite{Grumiller:2019ygj}.

\subsection{Conformal \texorpdfstring{AdS$_3$}{AdS3} geometries} 

As the next class of solutions, we consider geometries satisfying the conformal boundary conditions $\delta (\frac{q_{ab}}{\sqrt{-q}})=0$ and $\delta T=0$ with the flat boundary metric. In a suitable coordinate system, these conditions imply
\begin{equation}
  \frac{q_{ab}}{\sqrt{-q}} \d{}x^{a} \d{}x^{b} = - \d{}x^{+} \d{}x^{-}  \, , \qquad  T=0\, .
\end{equation}
The boundary metric $q_{ab}$ and the corresponding EMT $T_{ab}$ that satisfy the constraint equations are then given by 
\begin{equation}\label{q-T-conformal-infty}
    \begin{split}
    {q_{ab}} \d{}x^{a} \d{}x^{b}   &=  - h_+(x^+) h_-(x^-)\d{}x^{+} \d{}x^{-} \, ,\\
        {T}_{\pm \pm} =  -& \frac{c}{12 \pi\ell^2} h_{\pm}^2 \, , \qquad  {T}_{+-} = 0\, ,
    \end{split}
\end{equation}
where $h_{+}(x^+)$ and $h_{-}(x^-)$ are two arbitrary functions. If the boundary is taken to be cylindrical, these functions must be periodic: $h_\pm(x^\pm+2\pi)=h_\pm(x^\pm)$. Note that for this case  T$\bar{\text{T}}$ operator is a constant: $\ttbarb= \frac{c^2}{18\pi^2\ell^4}$. 

Using \eqref{h-coord-trans}, the full bulk line element compatible with conformal boundary conditions
\begin{equation}\label{conf-soln}
        \d{}s^2 = \ell^2 \frac{\d{}r^2}{r^2} - \left( \frac{r}{\ell}\, h_{+} \d{}x^+ - \frac{\ell}{r} {h_-} \d{}x^-\right)\left( \frac{r}{\ell}\, h_{-} \d{}x^- - \frac{\ell}{r} {h_+} \d{}x^+\right)\, ,
\end{equation}
with 
\begin{equation}
    \sqrt{-h}= \frac{1}{2} \Big(\frac{r^2}{\ell^2}-\frac{\ell^2}{r^2} \Big) h_{+} h_{-}\, .
\end{equation}
This solution was previously discussed in \cite{Afshar:2017okz, Adami:2020uwd}. The rBY-EMT is also given by
\begin{equation}\label{calT-conformal}
    \sqrt{-h}\, \mathcal{T}^{\pm\pm} = -\frac{c r^2}{6 \pi (r^4-\ell^4)} \frac{h_{\mp}}{h_{\pm}}\, , \qquad \sqrt{-h}\, \mathcal{T}^{+-} = - \frac{c \ell^2}{6\pi (r^4-\ell^4)}\, .
\end{equation}
From this, we find
\begin{equation}\label{r-geom-data-cbc}
    \begin{split}
    \mathcal{T} = -\frac{c \ell^2}{3\pi(r^4-\ell^4)} \, , \qquad \Ottbar= \frac{c^2}{18 \pi^2 (r^4-\ell^4)}\, .
    \end{split}
\end{equation}
Therefore, while ${\cal T}, \Ottbar$ are nonvanishing and have nontrivial $r$ dependence, $\delta {\cal T}, \delta \Ottbar$ are zero. One should also note that $\delta(\frac{\sqrt{-h}}{\sqrt{-q}})=0$.

We now draw attention to an important subtlety. Recall that we established that the conformal symplectic potential does not flow \eqref{non-flow-N-C}. Here, we revisit this point in more detail. From  \eqref{r-geom-data-cbc}, we obtain $\delta \mathcal{T} = 0$, which satisfies the first requirement for conformal boundary conditions. However, \eqref{conf-soln} shows that the conformal metric at a finite distance has a non-vanishing variation, $\delta \hat{h}_{ab} \neq 0$. To see this more explicitly, we note that
\begin{equation}
    \frac{q_{ab}}{\sqrt{-q}} = \frac{r^4+3 \ell^4}{r^{4}-\ell^4} \frac{h_{ab}}{\sqrt{-h}} + \frac{12\pi \ell^2}{c} \frac{\mathcal{T}_{ab}}{\sqrt{-h}}\,.
\end{equation}
From the conformal boundary condition $\delta \left( \frac{q_{ab}}{\sqrt{-q}} \right)=0$, we recover the flow of the conformal boundary condition
\begin{equation}\label{conformal-bc-flow}
\delta \left(\frac{h_{ab}}{\sqrt{-h}} \right) = - \frac{12\pi \ell^2}{c}  \frac{r^{4}-\ell^4}{r^4+3 \ell^4}  \delta \left(\frac{\mathcal{T}_{ab}}{\sqrt{-h}} \right), \, \qquad \delta {\cal T}=0.
\end{equation}
Note that since $\sqrt{-h}\sim r^2, {\cal T}^{ab}\sim {\cal O}(1)$ at large $r$, the above reduces to standard conformal boundary conditions at large $r$. 
In terms of the more general class of boundary conditions discussed in  \eqref{Interpolating-class}, this corresponds to  $X={\cal T}, Y=1, {\cal F}(-h)=-\frac{c}{12\pi\ell^2}\frac{1}{\sqrt{-h}}-\frac{X}{2}$.\footnote{Note that \eqref{conformal-bc-flow} may also be written as $\delta \left({\sqrt{-h}}{\mathcal{T}^{ab}} \right)=-\frac{c}{12\pi\ell^2}\delta h^{ab}
$. It is instructive to compare this with the boundary condition written in footnote \ref{footnote-9}.}
Nevertheless, the condition $\mathcal{T}^{ab} \delta \hat{h}_{ab} = 0$ still holds. This shows that while the boundary conditions at asymptotic infinity ($\Sigma$) and at a finite radial cutoff ($\Sigma_r$) differ in form, both remain consistent with vanishing of the conformal symplectic potential, $\Theta_{\text{\tiny{C}}}(\Sigma_r) =0= \Theta_{\text{\tiny{C}}}(\Sigma)$.

Despite this distinction, both the boundary conditions at infinity and at a finite cutoff lie within the class of conformal boundary conditions. As a result—unlike in the case of Dirichlet boundary conditions, which require a T$\bar{\text{T}}$-like deformation of the dual $2d$ QFT to account for radial evolution—no such deformation is needed for the geometries described in \eqref{conf-soln}. See also \cite{Adami:2020uwd} for related discussions.

\paragraph{Surface charges and their algebra.} As shown in  Appendix \ref{appen:conformal-charges}, the charge algebra is two copies of centerless Virasoro (Witt) algebras. 

\subsection{Conformal conjugate geometries} 
The conformal conjugate boundary condition, first introduced in \cite{Parvizi:2025shq, Parvizi:2025wsg} as the conjugate counterpart of the well-known conformal boundary condition, is defined as follows:
\begin{equation}
    \delta \sqrt{-q}=0\, , \qquad  \delta \left(T^{ab}-\frac{T}{2} q^{ab}\right)=0\, .
\end{equation}
The Einstein constraint equations \eqref{cons-trace} with this boundary condition do not admit a solution involving two arbitrary codimension-two functions. The reason is that the system imposes four independent conditions rather than three. Consequently, one can only obtain a solution with a single codimension-two free function
\begin{equation}
    \sqrt{-q} = 1\, , \qquad     T^{ab}-\frac{T}{2} q^{ab} = \frac{c}{12 \pi \ell^2} \begin{bmatrix}
0 & 0 \\
0 & 1
\end{bmatrix}\, .
\end{equation}
This boundary condition yields the following solution
\begin{equation}
    \d{}s_q^{2} = 2 \d{}t \d{}\phi + g(\phi) \d{} \phi^2\, ,
\end{equation}
where $g(\phi)$ is an arbitrary function. Then, the bulk line element is given by
\begin{equation}
        \d{}s^2 = \ell^2 \frac{\d{}r^2}{r^2} + 2\frac{r^2}{\ell^2} \d{}t \d{} \phi + \left( \frac{r^2}{\ell^2}g(\phi) - 1 \right)\d{}\phi^2\, .
\end{equation}
This is indeed the Neumann AdS$_3$ geometry \eqref{metric-N-bc-Sigma} with $f(\phi)=1$.

\subsection{Generic \texorpdfstring{$w$}{w} case}

In \eqref{bulk-action-W-3} we introduced a more general one-parameter family of $3d$ gravity actions. For this action the symplectic potential is given as (see section 6.1 of \cite{Parvizi:2025wsg} for more detailed discussions, where $w$ was denoted as $k$)
\begin{equation}\label{w-Theta}
    \begin{split}
        \Theta_{w}(\Sigma_r) &= -\frac{1}{2} \int_{\Sigma_r} \left[ (1-w) \sqrt{-h}\, \mathcal{T}^{ab}\ \delta h_{ab} -w\ \delta (\sqrt{-h} {\cal T}^{ab})\ {h}_{ab} \right]\, ,\\
        &= -\frac{1}{2} \int_{\Sigma_r} \ \sqrt{-\tilde{h}}\, \tilde{\mathcal{T}}^{ab}\ \delta \tilde{h}_{ab}
    \end{split}
\end{equation}
where 
\begin{equation}
    \tilde{h}_{ab}= \Omega^{-1} h_{ab},\qquad \tilde{\mathcal{T}}^{ab}=\Omega^{2} (\mathcal{T}^{ab}-\frac{w}{2}\mathcal{T} h^{ab}),\qquad \Omega={\cal T}^{\frac{w}{1-w}}.
\end{equation}
The above interpolates between Dirichlet ($w=0$) and Neumann ($w=1$) boundary conditions. Since we have already discussed in detail the $w=0,1$ cases, here we focus on  $w\neq 0,1$. Except for the $w=1$ case for which ${\cal T}$ vanishes and the above description in terms of tilde quantities is ill-defined, in generic $w$ case, one can use the tilde-frame to construct explicit solutions. 

Here we focus on the Dirichlet boundary conditions in the tilde-frame: $\delta \tilde{h}_{ab}=0$. Thus, the solution space may be constructed based on the Ba\~nados geometries discussed in section \ref{sec:Banados}, explicitly, 
\begin{equation}
    \d{}s^2_w= {\cal T}^{\frac{w}{1-w}}\ \d{}s^2_{\text{\tiny{Ba\~nados}}},
\end{equation}
where ${\cal T}$ is given in \eqref{calT-Babandos} and $\d{}s^2_{\text{\tiny{Ba\~nados}}}$ is the metric in \eqref{Banados-3D}.

\subsection{Black flowers}\label{sec:blackflower}

As the next example, we consider the black flower solution \cite{Afshar:2016wfy}
\begin{equation}\label{blackflower-metric}
    \begin{split}
        \d{}s^2 &= -  \frac{r}{\ell} f \d{}t^2 + 2 \d{}t \d{}r + 2 r f \, J_- \d{}t \d{} \phi - 2\ell J_- \d{}r \d{}\phi + \ell^2\left[ \frac{r}{\ell} f (J_+^2-J_-^2) +\frac{1}{4} J_+^2 \right] \d{}\phi^2,
    \end{split}
\end{equation}
where  $\phi\equiv \phi+2\pi$ and, $J_+(\phi)$ and $J_-(\phi)$ are two arbitrary periodic functions of $\phi$ and $f$ is given by $f(r)=1+\frac{r}{\ell}$. Note that this is slightly different than the black flower solution discussed in \cite{Afshar:2016wfy}: here $J_\pm (\phi)$ are not necessarily of the form of $\partial_\phi \varphi_\pm$. We will discuss this difference at the end of this subsection. The flower line element \eqref{blackflower-metric} is not in the standard FG gauge. The corresponding lapse function and shift vector \eqref{metric} are
\begin{equation}
    N^{2} = \frac{\ell}{r f} = - U^{t} \, , \qquad U^{\phi} = 0\, ,
\end{equation}
and the induced metric is given by
\begin{equation}
    \begin{split}
        \d{}s_{h}^2 =  & - \frac{r}{\ell} f \d{}t^2  + 2 r f J_- \d{}t \d{} \phi + \ell^2 \left[ \frac{r f}{\ell}(J_+^2-J_-^2) +\frac{ J_+^2 }{4}\right] \d{}\phi^2\, ,
    \end{split}
\end{equation}
with determinant
\begin{equation}
    \sqrt{-h} = \frac{r^2}{\ell}\Big(1 + \frac{\ell }{2r} \Big)\sqrt{1+\frac{\ell }{r}}\, J_{+}\, .
\end{equation}
The boundary conformal induced metric and asymptotic boundary EMT are, 
\begin{equation}
\begin{split}
    \d{}s^{2}_{q} &= -  \d{}t^2 + 2 \ell J_- \d{}t \d{}\phi +  \ell^2 (J_+^2-J_-^2) \d{}\phi^2\, ,  \\  T_{ab} \d{}x^{a} \d{}x^{b}  &= \frac{c}{96 \pi \ell^2} \left[- \d{}t^2 + 2\ell J_- \d{}t \d{}\phi - \ell^2 (J_+^2+J_-^2) \d{}\phi^2 \right]\, ,
\end{split}
\end{equation}
with $\sqrt{-q} = \ell\, J_{+}$. 
In this case, we have
\begin{equation}
    T=0\, , \qquad \ttbarb = \frac{\ell^2}{32}\,.
\end{equation}
As we see, the above is like the conformal boundary condition case. 
The rBY-EMT at a finite distance $r$ is written in terms of boundary data as follows
\begin{equation}
    \mathcal{T}_{ab} = \mathcal{F}_{1}(r) q_{ab} + \mathcal{F}_2(r) T_{ab}\, ,
\end{equation}
with 
\begin{equation}
   \begin{split}
       &  \mathcal{F}_{1}(r) = \frac{c}{192 \pi \ell^4} \frac{-32 r^{2}(r+\ell)^2+[8r^2+8r \ell+\ell^2][2(2r +\ell) \sqrt{r(r+\ell)}-\ell^2]}{(2r +\ell) \sqrt{r(r+\ell)}}\, , \\
       &  \mathcal{F}_{2}(r) = \frac{1}{2}\left[ -2 + \sqrt{\frac{r}{r+\ell}} + \sqrt{\frac{r+\ell}{r}} + \frac{4 \sqrt{r(r+\ell)}}{2r+\ell} \right]\, ,
   \end{split}
\end{equation}
with the following asymptotic behavior 
\begin{equation}
    \mathcal{F}_{1}(r) = - \frac{c}{512 \pi r^2} + \mathcal{O}(r^{-3})\, , \qquad \mathcal{F}_{2}(r) = 1 + \mathcal{O}(r^{-4})\, .
\end{equation}
Similarly, the induced metric on $\Sigma_r$, $h_{ab}$, can be written as follows
\begin{equation}
    h_{ab} = \left( \frac{1}{8} + \frac{r f}{\ell} \right) q_{ab} - \frac{12 \pi \ell^2}{c} T_{ab}\, .
\end{equation}
The trace of the rBY-EMT at a finite distance is given by
\begin{equation}
    \mathcal{T} = -\frac{6\pi\ell^2}{c}\Ottbar = \frac{c}{12 \pi \ell} \left[\frac{2}{\ell} - \frac{(1+\frac{8r}{\ell}f)(\frac{r}{\ell} + f)}{2\sqrt{ \ell r f} (1+\frac{4r}{\ell}f)}\right]\, .
\end{equation}
As we see, like the conformal boundary condition case, $\delta(\frac{\sqrt{-h}}{\sqrt{-q}})=0$ and $\delta{\cal T}=0=\delta\Ottbar$. 
\paragraph{Boundary condition.} 
The asymptotic Dirichlet symplectic potential is given by
\begin{equation}
    \Theta_{\text{\tiny{D}}}(\Sigma) = -\frac{1}{2} \int_{\Sigma} \sqrt{-q} T^{ab} \delta q_{ab} = \frac{1}{8\kappa} \delta \int_{\Sigma}  J_+ = \frac{1}{8\kappa \ell} \delta \int_{\Sigma} \sqrt{-q}\, .
\end{equation}
The above does not vanish in general, and ensuring a well-defined action principle requires considering two possibilities: 
\begin{enumerate}
\item 
 $J_+(\phi)= \partial_{\phi} \varphi_+(\phi)$. This is the case considered in \cite{Afshar:2016wfy} and the Dirichlet symplectic potential vanishes for the black flower solution.
\item Extended Black Flower: One may add a term to $\Theta_{\text{\tiny{D}}}(\Sigma)$ so that the new symplectic potential vanishes for the black flower solution space. The new symplectic potential is
\begin{equation}
     \Theta_{\text{\tiny{Extended flower}}}(\Sigma) \;=\; -\tfrac{1}{2} \int_{\Sigma} \sqrt{-q}\, \mathbb{T}^{ab}\, \delta q_{ab} \;=\; 0\, , \qquad \mathbb{T}^{ab} := T^{ab} + \frac{1}{8 \ell \kappa }\, q^{ab}\, .
\end{equation}
This above is a special case of \eqref{bulk-action-W-2} with $w=0, b=-1+\frac18$. The new EMT has a simple form
\begin{equation}
    \mathbb{T}^{ab} = -\frac{c}{48 \pi \ell^2}  \begin{bmatrix}
1 & 0 \\
0 & 0
\end{bmatrix}\, .
\end{equation}
To write the explicit form of the boundary conditions associated with the black flower, we rewrite the black flower symplectic potential as follows
\begin{equation}
     \Theta_{\text{\tiny{Extended flower}}}(\Sigma) \;=\; -\tfrac{1}{2} \int_{\Sigma} \sqrt{-q}\, \left( \delta \mathbb{T} - q_{ab}\, \delta \mathbb{T}^{ab} \right)\;=\; 0\, .
\end{equation}
Therefore, one can consider the black flower as a solution to the Einstein field equation with the following 
\begin{equation}
    \delta \mathbb{T} = 0\, , \qquad \delta \mathbb{T}^{ab}=0\, .
\end{equation}
In terms of the original data, the black flower corresponds to the following mixed boundary condition,
\begin{equation}\label{BF-bc}
    \delta T=0\, , \qquad \delta \Big(T^{ab} + \frac{c}{96 \pi \ell^2}\, q^{ab} \Big)=0\, .
\end{equation}
\end{enumerate}
\paragraph{Boundary condition flow.}
Regarding the black flower boundary condition flow, we observe that the mixed combination appearing in the black flower boundary condition \eqref{BF-bc} can be expressed in terms of the induced metric at finite distance and the rBY energy–momentum tensor as follows
\begin{equation}
    T^{ab} + \frac{c}{96 \pi \ell^2}\, q^{ab} = \mathcal{H}_{1}(r) \mathcal{T}^{ab} + \mathcal{H}_{2}(r) h^{ab}\, , 
\end{equation}
where $\mathcal{H}_{1}(r)$ and $\mathcal{H}_{2}(r)$ are given by
\begin{equation}
   \begin{split}
       & \mathcal{H}_{1}(r) = \frac{(\ell + 2r) [r(r+\ell)]^{3/2}}{2\ell^4}\, , \\
       & \mathcal{H}_{2}(r) = \frac{c}{48 \pi \ell^6 } r (r+\ell)(2r+\ell)\left(\sqrt{r+\ell}-\sqrt{r} \right)^2\, .
   \end{split}
\end{equation}
Therefore, the asymptotic boundary condition induces the following finite-$r$ boundary condition
\begin{equation}\label{BF-bc-finite}
    \delta \left[ \mathcal{T}^{ab} + \frac{c}{24\pi \ell^2}\frac{\big(\sqrt{r+\ell}-\sqrt{r} \big)^2}{\sqrt{r(r+\ell)}} h^{ab} \right] =0\, ,\qquad \delta {\cal T}=0\, .
\end{equation}
\paragraph{Boundary action.}
To derive the deformation flow equation for the black flower case we  first need to identify the boundary term responsible for generating the black flower boundary condition
\begin{equation}\label{Theta-BF-finite-1}
    \Theta_{\text{\tiny{BF}}}(\Sigma_r) = \Theta_{\text{\tiny{{D}}}}(\Sigma_r) - \delta \int_{\Sigma_r} \mathcal{F}(r)\, \sqrt{-h}\, , \qquad   \mathcal{F}(r) :=  \frac{c}{24\pi \ell^2}\frac{\big(\sqrt{r+\ell}-\sqrt{r} \big)^2}{\sqrt{r(r+\ell)}}\, .
\end{equation}
Then, the black flower symplectic potential is written as
\begin{equation}
    \begin{split}
        \Theta_{\text{\tiny{BF}}}(\Sigma_r)& =  -\frac{1}{2} \int_{\Sigma_r} \sqrt{-h}\, \left[\mathcal{T}^{ab} + \mathcal{F}(r) h^{ab} \right] \delta h_{ab} \\
        & = \frac{1}{2} \int_{\Sigma_r} \left[ \sqrt{-h}\, h_{ab}\, \delta \left( \mathcal{T}^{ab} + \mathcal{F}(r) h^{ab}  \right) - \sqrt{-h}\,  \delta \mathcal{T} \right]\, .
    \end{split}
\end{equation}
It is straightforward to see that the above is consistent with the black flower boundary condition \eqref{BF-bc-finite}. Finally, using \eqref{Theta-BF-finite-1}, we obtain the deformation flow equation
\begin{equation}
     {S}^{\text{\tiny{BF}}}_{\text{bdry}}[\Sigma_r] =  {S}^{\text{\tiny{D}}}_{\text{bdry}}[\Sigma_r] - \int_{\Sigma_r} \mathcal{F}(r)\, \sqrt{-h}\, .
\end{equation}
By using \eqref{Integrated-effective-action-D}, we find the explicit form boundary action as follows
\begin{equation*}
 \hspace*{-6mm}
 \inbox{
      \hspace*{-5mm}  {S}^{\text{\tiny{BF}}}_{\text{bdry}}[\Sigma_r]  = {S}^{\text{\tiny{D}}}_{\text{bdry}}[\Sigma] + \frac{c \text{E} }{12} \ln \left( \frac{r}{r_\infty} \right) + \frac{3\pi\ell^4}{c r^2} \int \text{d}^2x \sqrt{-q} \ttbarb  
        - \frac{c}{48\pi \ell^2} \frac{2r+\ell}{(\sqrt{r+\ell} +\sqrt{r})^2} \int \text{d}^2x \sqrt{-q} 
  \hspace*{-4mm}}
\end{equation*}
In the large $r$ limit, we recover 
\begin{equation}
    {S}^{\text{\tiny{BF}}}_{\text{bdry}}[\Sigma] = {S}^{\text{\tiny{D}}}_{\text{bdry}}[\Sigma] - \frac{c}{96 \pi \ell^2} \int_{\Sigma} \sqrt{-q}\, .
\end{equation}

We close this part noting that the surface charge algebra for this case, like the conformal boundary condition case, is the centerless Witt algebra, cf. Appendix \ref{appen:blackflower-charges}.

\section{Solutions with non-covariant boundary conditions}\label{sec:noncovariant-bc-solnspace}

In this section we construct and study some examples of solution spaces with non-covariant boundary conditions within the general formulation developed in section \ref{sec:non-covariant-bc}. Here we discuss three examples, the first one is an extension of the the Compere-Song-Strominger solutions \cite{Compere:2013bya} while the other two are new kind of solutions not considered in the literature before. 

\subsection{Compere-Song-Strominger geometries}\label{sec:CSS}

As the first example of solutions with non-covariant boundary conditions we consider  the case that is an extension of the Compere-Song-Strominger (CSS) geometries \cite{Compere:2013bya}. We relax the demanding fixing of a combination of $q_{ab}$, $\sqrt{-q}$, ${T}$, and $\sqrt{-q}\, {T}^{ab}$.  The CSS boundary conditions are given by $\delta{\sqrt{-q}}=0$, and the metric has a fixed null direction $t^a$. $T_{ab}$ has a vanishing variation, except for components along the null direction $t^a$, such that all in all, $T^{ab}\delta q_{ab}=0$. 

In terms of the notations in section \ref{sec:non-covariant-bc} and recalling \eqref{non-covariant-bc}, a slightly extended version of the CSS boundary condition can be represented as\footnote{Note that in this section, we apply the analysis of section \ref{sec:non-covariant-bc} at infinity with canonical pair $\{q_{ab}, T^{ab}\}$.}
\begin{equation}\label{CSS-bc}
   \begin{split}
    t\cdot t :=0\, , \qquad
   t \cdot k :=-1\, , \qquad  \delta(\sqrt{-q}\, Z) = \delta \Delta\, ,
   \end{split}
\end{equation}
where $\Delta$ is a constant and by $t\cdot t :=0$ we mean $t\cdot t =0, \delta(t\cdot t)=0$ and similarly  $t\cdot k:=-1$ means $t\cdot k=-1, \delta(t\cdot k)=0$. That is, $t$ is a null vector field orthogonal to the vector field $k$ and we adopt the following representations for the two background vector fields $t^{a}$ and $k^{a}$,
\begin{equation}\begin{split}
t^{a}\partial_{a}= \partial_{t}\, , \qquad k^{a}\partial_{a} = \partial_{\phi} \, , \quad &\text{with } \quad \delta t^{a} =0\, , \qquad \delta k^{a} = 0\, \\
    t\cdot t := 0\, &, \qquad  t \cdot k :=-1\, .
\end{split}\end{equation}
The most general form of the boundary metric compatible with the above boundary condition is 
\begin{equation}
     q_{ab} = -(k \cdot k) t_{a} t_{b} + t_{a} k_{b} + t_{b} k_{a}\, ,
\end{equation}
which yields $\sqrt{-q}=1$. From  \eqref{X-Y-Z} and the above boundary conditions, we obtain
\begin{equation}\label{ZTttDelta}
  \sqrt{-q}\,  Z = T_{tt} = \Delta\, .
\end{equation}
We consider the AdS boundary to be flat; consequently, the trace anomaly equation yields
\begin{equation}
    T = -2 \mathrm{T}_{tk} + (k\cdot k) \mathrm{T}_{tt}=0\, .
\end{equation}
Hence, the most general form of the boundary energy-momentum tensor is given by
\begin{equation}
    T_{ab} = \mathrm{T}_{kk} t_{a} t_{b} - \frac{\Delta}{2}(k\cdot k) (t_{a}k_{b}+k_{a}t_{b}) + \Delta\, k_{a}k_{b}\, .
\end{equation}
We are thus left with two arbitrary functions, $k\cdot k$ and $\mathrm{T}_{kk}$. The final step is to impose the conservation equation, $\nabla_a T^{ab} = 0$, which constrains these functions. Specifically, this equation expresses the two arbitrary codimension-one functions in terms of two arbitrary codimension-two functions. The resulting expressions are
\begin{equation}
    \begin{split}
    &{q_{ab}} \d{}x^{a} \d{}x^{b}  =  2\d{}t \d{}\phi + J(\phi) \d{}\phi^2 \, ,\\
   &{T}_{ab} \d{}x^{a} \d{}x^{b} =  \Delta \d{}t^2+ \Delta J(\phi) \d{}t \d{}\phi  + [L(\phi)+\Delta J(\phi)^2]\d{}\phi^2\, ,
    \end{split}
\end{equation}
where $\phi\equiv \phi+2\pi$ and $J(\phi), L(\phi)$ are  taken to be $2\pi$ periodic arbitrary functions. Inserting the above into our general formula \eqref{h-coord-trans}, we obtain the corresponding $3d$ solution is \cite{Compere:2013bya},
\begin{equation}\label{CSS-soln}
        \begin{split}
            \d{}s^2 & = \ell^2 \frac{\d{}r^2}{r^2} - \frac{12 \pi \ell^2 \Delta}{c} \d{}t^2 + 2\left[ \frac{r^2}{\ell^2}-\frac{6\pi \ell^2 \Delta}{c} J + \Delta \left(\frac{3 \pi \ell^3}{c r} \right)^2 (4 L +3 \Delta J^2) \right] \d{}t\d{}\phi \\
            & +\frac{ c r^2 - 3\pi \ell^4 \Delta J }{c^2 \ell^2 r^2}\left( c r^2 J - 9\pi \ell^4 \Delta J^2- 12 \pi \ell^4 L \right) \d{}\phi^2\, ,
        \end{split}
    \end{equation}
and the associated rBY-EMT is
\begin{equation}
    \mathcal{T}_{ab} = -\frac{9 \pi \ell^4 \ttbarb}{c r^2} \frac{1 + \frac{6\pi^2 \ell^8}{c^2 r^4} \ttbarb}{ 1 - \frac{18\pi^2 \ell^8}{c^2 r^4}\ttbarb} q_{ab} + \left( -3 + \frac{4}{1 - \frac{18 \pi^2 \ell^8 }{c^2 r^4}\ttbarb} \right) T_{ab}\, .
\end{equation}
In this case, we have
\begin{equation}
    \begin{split}   
    T=0\, , \quad &\quad \ttbarb = \frac{\Delta}{2} \left( 4 L+3 \Delta J^2 \right)\, , \\
    \sqrt{-h}= \frac{r^2}{\ell^2}-\left(\frac{3\pi \ell^3}{c r}\right)^2\Delta \left(4 L+3\Delta J^2\right), &\quad  
       \sqrt{-h}\  \Ottbar = \frac{\ell^2}{r^2} \ttbarb\, ,  \quad \mathcal{T} = -\frac{6\pi \ell^2}{c}\Ottbar\, . 
    \end{split}
\end{equation}
The  Dirichlet symplectic potential for the above solution is of the form
\begin{equation}
    \Theta_{\text{\tiny{D}}}(\Sigma) = -\frac{1}{2} \int_{\Sigma} \sqrt{-q} T^{ab} \delta q_{ab} = -\frac{\Delta}{2}\delta \Big(  \int_{\Sigma} J \Big)\, .
\end{equation}
As we see it does not vanish. There are two options for having a well-defined action principle: 
\begin{enumerate}
    \item The choice made in  \cite{Compere:2013bya}, $J(\phi) = \partial_{\phi}P(\phi)$ where $P(\phi)$ is a periodic function.  In this case  $\Theta_{\text{\tiny{D}}}(\Sigma)=0$. However, in this case $\delta q_{ab} \neq 0$, so it is different from the standard Dirichlet boundary condition. 

\item Adding an extra boundary term to the GHY term as $W=\frac{\Delta  J}{2}$ and keep $J_0:=\int_0^{2\pi} \d{}\phi J(\phi)\neq 0$. This is an extension of the CSS case along the discussions in section \ref{sec:bdry-vs-corner} and the addition of a corner term. The symplectic potential then becomes,  
\begin{equation}
   \begin{split}
       \Theta_{\text{\tiny{Extended CSS}}}[\Sigma] & =  -\frac{1}{2} \int_{\Sigma} \sqrt{-q}\, T^{ab} \delta q_{ab} +  \frac{1}{2} \int_{\Sigma} \sqrt{-q}\, \mathrm{T}_{tt}\, \delta(k \cdot k) \\
       & = - \frac{1}{2} \int_{\Sigma} \sqrt{-q}\, \left[ \mathrm{T}_{kk} \delta(t \cdot t) - 2 \mathrm{T}_{tk} \delta (t \cdot k) \right] =0\, .
   \end{split}
\end{equation}
This symplectic potential is manifestly compatible with CSS boundary conditions \eqref{CSS-bc} and clearly breaks the covariance at the boundary $\Sigma$. 
\item If we allow $\Delta, J_0$, while constant on spacetime, to have nonzero variations on the solution space, $\Delta, J_0$ appear as canonical conjugates on the solution space, $[\Delta, J_0]={2i}$. In comparison to the CSS $u(1)$ Kac-Moody boundary symmetries algebra \cite{Compere:2013bya}, now $J_0$ is not the center of charge algebra, and $\Delta$ also appears in the algebra, see Appendix \ref{appen:CSS-charges}.

\end{enumerate}

\paragraph{Boundary condition flow.}
Recall \eqref{qab-again},
\begin{equation}
  q_{ab} = \frac{\ell^2}{r^2}\left( 1 - \frac{3 \pi \ell^2}{c} \mathcal{T} \right)\left[( 1 - \frac{12 \pi \ell^2}{c} \mathcal{T} )  h_{ab} + \frac{12 \pi \ell^2}{c } \mathcal{T}_{ab}\right]\, .
\end{equation}
By contracting both sides of this equation with $t^{a}t^{b}$, $t^{a}k^{b}$, and $k^{a}k^{b}$, we obtain, respectively,
\begin{equation}\label{norm-vectors}
    \begin{split}
        & (t\cdot t)_{q} = \frac{\ell^2}{r^2}\left( 1 - \frac{3 \pi \ell^2}{c} \mathcal{T} \right)\left[( 1 - \frac{12 \pi \ell^2}{c} \mathcal{T} )  (t \cdot t)_{h} + \frac{12 \pi \ell^2}{c } \mathcal{T}_{tt}\right]\, , \\
        & (t\cdot k)_{q} = \frac{\ell^2}{r^2}\left( 1 - \frac{3 \pi \ell^2}{c} \mathcal{T} \right)\left[( 1 - \frac{12 \pi \ell^2}{c} \mathcal{T} )  (t \cdot k)_{h} + \frac{12 \pi \ell^2}{c } \mathcal{T}_{tk}\right]\, , \\
        &(k\cdot k)_{q} = \frac{\ell^2}{r^2}\left( 1 - \frac{3 \pi \ell^2}{c} \mathcal{T} \right)\left[( 1 - \frac{12 \pi \ell^2}{c} \mathcal{T} )  (k \cdot k)_{h} + \frac{12 \pi \ell^2}{c } \mathcal{T}_{kk}\right]\, , \\
    \end{split}
\end{equation}
where the subscripts $q$ and $h$ indicate that the contractions are performed with the metrics $q_{ab}$ and $h_{ab}$, respectively.

We are now ready to extract the CSS boundary condition flow. The first two CSS boundary conditions at the AdS boundary \eqref{CSS-bc} induce the following mixed, non-covariant boundary condition at a finite radial cutoff:
\begin{equation}
    \begin{split}
        & \delta \left[ ( 1 - \frac{3 \pi \ell^2}{c} \mathcal{T} )\left(( 1 - \frac{12 \pi \ell^2}{c} \mathcal{T} )  (t \cdot t)_{h} + \frac{12 \pi \ell^2}{c} \mathcal{T}_{tt} \right)\right]=0\, , \\
        & \delta \left[ ( 1 - \frac{3 \pi \ell^2}{c} \mathcal{T} )\left(( 1 - \frac{12 \pi \ell^2}{c} \mathcal{T} )  (t \cdot k)_{h} + \frac{12 \pi \ell^2}{c} \mathcal{T}_{tk} \right)\right]=0\, .
    \end{split}
\end{equation}
 Now we consider the flow of the third boundary condition \eqref{CSS-bc} or \eqref{ZTttDelta}.
We need to express the asymptotic variables in terms of finite-distance quantities. As a first step, we write the boundary EMT in terms of the finite-distance variables $h_{ab}$ and $\mathcal{T}_{ab}$
\begin{equation}
    T_{ab} = \frac{3}{2} \mathcal{T} \left( -1 + \frac{4\pi \ell^2}{c}\mathcal{T} \right) h_{ab} + \left( 1 - \frac{6\pi \ell^2}{c} \mathcal{T} \right) \mathcal{T}_{ab}\, .
\end{equation}
This equation suffices to read 
\begin{equation}\label{asymptotic-EMT-component}
   \begin{split}
        T_{kk} = & \frac{3}{2} \mathcal{T} \left( -1 + \frac{4\pi \ell^2}{c}\mathcal{T} \right) (k \cdot k)_{h} + \left( 1 - \frac{6\pi \ell^2}{c} \mathcal{T} \right) \mathcal{T}_{kk}\, , \\
        T_{tk} = & \frac{3}{2} \mathcal{T} \left( -1 + \frac{4\pi \ell^2}{c}\mathcal{T} \right) (t \cdot k)_{h} + \left( 1 - \frac{6\pi \ell^2}{c} \mathcal{T} \right) \mathcal{T}_{tk}\, , \\
        T_{tt} = & \frac{3}{2} \mathcal{T} \left( -1 + \frac{4\pi \ell^2}{c}\mathcal{T} \right) (t \cdot t)_{h} + \left( 1 - \frac{6\pi \ell^2}{c} \mathcal{T} \right) \mathcal{T}_{tt}\, . \\
   \end{split}
\end{equation}
To complete, we relate determinants
\begin{equation}\label{det-CSS}
    \sqrt{-q} = \frac{\ell^2}{r^2} \sqrt{-h} \left( 1  - \frac{3 \pi \ell^2}{c } \mathcal{T}\right)\, .
\end{equation}
Thus, the final result is obtained by substituting \eqref{norm-vectors}, \eqref{asymptotic-EMT-component}, and \eqref{det-CSS} into \eqref{ZTttDelta}.
\paragraph{Boundary action.}
Finally, we discuss the finite cutoff boundary action associated with the CSS boundary conditions
\begin{equation}
  \inbox{  {S}^{\text{\tiny{CSS}}}_{\text{bdry}}[\Sigma_r] = {S}^{\text{\tiny{D}}}_{\text{bdry}}[\Sigma_r] - \int_{\Sigma_r} \sqrt{-h}\, Z\, (k\cdot k)_h \,, }
\end{equation}
where \( S^{\text{\tiny D}}_{\text{bdry}}[\Sigma_r] \) is defined in \eqref{Integrated-effective-action-D}, and \( Z \) is given by \eqref{X-Y-Z}. As is evident, the last term involves background vector fields and is therefore non-covariant. The non-covariance of the CSS boundary conditions is thus also reflected in the boundary action.

\subsection{Geometries with a fixed null direction}\label{sec:new-c-bc}
In this subsection, we present a new class of solutions for which, unlike the CSS case, $\delta{\sqrt{-q}}\neq 0$, but with $\delta(q_{ab}/{\sqrt{-q}})=0$, i.e. we fix the boundary metric up to a conformal factor. Explicitly, we consider the following boundary conditions
\begin{equation}\label{non-cov-bc-null}
    t \cdot t :=0\, , \qquad \delta (\sqrt{-q}\, Y)= \delta \tilde{\Delta}\, , \qquad \delta (\sqrt{-q}\, Z)=\delta \Delta\, ,
\end{equation}
where $t \cdot t :=0$ means $t \cdot t =0, \delta(t \cdot t)=0$, and $\Delta$ and $\tilde{\Delta}$ are two arbitrary constant and $Y$ and $Z$ have been defined in \eqref{X-Y-Z}. The following solution satisfies all the above boundary conditions
{
\begin{equation}
    \begin{split}
        & \d{}s_{q}^2 = -(t\cdot k) \left( -2 \d{}t \d{}\phi + \frac{2\tilde{\Delta}}{\Delta} \d{}\phi^2 \right)\, , \\
        & {T}_{ab} \d{}x^a \d{}x^b = \Delta\, (t\cdot k) \d{}t^2 - 2 (t\cdot k)\tilde{\Delta} \d{}t\, \d{}\phi +F\d{}\phi^2\, .
    \end{split}
\end{equation}
By imposing $\nabla_a T^{ab}=0$, we find
\begin{equation}
    t\cdot k = -f(t-\frac{\tilde{\Delta}}{\Delta}\phi)\, , \qquad F = g(\phi) - \frac{\tilde{\Delta}^2}{\Delta}  f(t-\frac{\tilde{\Delta}}{\Delta}\phi)\, .
\end{equation}
}
This solution has the following properties
\begin{equation}
    T=0\, , \qquad \ttbarb = -\frac{2 g}{f}\Delta\, .
\end{equation}
The induced metric and the rBY-EMT at a finite distance are given by
\begin{equation}
    \begin{split}
        h_{ab} = & \left( \frac{r^2}{\ell^2} + \frac{18 \pi^2 \ell^6 \ttbarb}{c^2 r^2} \right) q_{ab} - \frac{12 \pi \ell^2}{c} T_{ab}\, , \\
        \mathcal{T}_{ab} = & - \frac{9 \pi \ell^4\ttbarb (c^2 r^4 + 6 \pi^2 \ell^8 \ttbarb)}{c r^2 (c^2 r^4 - 18 \pi^2 \ell^8 \ttbarb) } q_{ab} + \frac{c^2 r^4 + 54 \pi^2 \ell^8 \ttbarb}{c^2 r^4 - 18 \pi^2 \ell^8 \ttbarb} T_{ab}\, ,
    \end{split}
\end{equation}
with
\begin{equation}
    \mathcal{T} =  \frac{12 \pi \ell^6\, c\, \Delta\, g}{c^2 r^4 f + 36 \pi^2 \ell^8 \Delta\, g}\, , \qquad \Ottbar = - \frac{6\pi \ell^2}{c}\mathcal{T}
\end{equation} 
\paragraph{Boundary action.}
The boundary action corresponding to the boundary condition \eqref{non-cov-bc-null} takes the form
\begin{equation}\label{bdry-action-non-cov-bc-null}
  \inbox{  {S}^{\text{\tiny{CSS}}}_{\text{bdry}}[\Sigma_r] = {S}^{\text{\tiny{D}}}_{\text{bdry}}[\Sigma_r] - \int_{\Sigma_r} \sqrt{-h} \left[ 2 Y\, (t \cdot k)_h + Z\, (k \cdot k)_h \right]\,, }
\end{equation}
where \( S^{\text{\tiny D}}_{\text{bdry}}[\Sigma_r] \) is defined in \eqref{Integrated-effective-action-D}, and \( X \) and \( Z \) are given by \eqref{X-Y-Z}. The non-covariant nature of the boundary conditions \eqref{non-cov-bc-null} manifests itself in the appearance of the last term, which explicitly breaks covariance in the boundary action. As the final comment, the charge algebra associated with this class of solutions two copies of Virasoro algebras at Brown-Henneaux central charge plus a ``center of mass'' Heisenberg algebra; see appendix \ref{appen:charge-section5.2} for the details of derivation. 
\subsection{A new non-covariant boundary condition}\label{sec:new-non-covariant}
In this subsection, we introduce a new non-covariant boundary condition. Let us take the following boundary condition at infinity
\begin{equation}
    t \cdot t :=-1\, , \qquad \delta (\sqrt{-q}\, Y)= \delta \chi\, , \qquad \sqrt{-q}\, Z : =0\, .
\end{equation}
The most general solution with these boundary conditions is given by
\begin{subequations}
    \begin{align}
        \d{}s_q^2 = & - \d{}t^2 - 2g(\phi) \d{}t \d{}\phi + [f(\phi)^2- g(\phi)^2] \d{}\phi^2\, , \\
        T_{ab} =  & -2 \chi f(\phi) \d{}t \d{}\phi -2 \chi f(\phi)\, g(\phi) \d{}\phi^2\, ,
    \end{align}
\end{subequations}
with the following properties
\begin{equation}
    T=0\, , \qquad \ttbarb = -2 \chi^2\, .
\end{equation}
The induced metric and rBY-EMT at a finite distance also written as follows\footnote{We note that, despite the similarity, this solution is different than the black flower discussed in section \ref{sec:blackflower}; this solution is in the FG gauge whereas \eqref{blackflower-metric} is not. The surface charge analysis also uncovers their difference. }
\begin{subequations}
    \begin{align}
        h_{ab} = & \left( \frac{r^2}{\ell^2} - \frac{36\pi^2 \ell^6 \chi^2 }{c^2 r^2} \right) q_{ab} - \frac{12\pi \ell^2}{c}T_{ab}\, , \\
        \mathcal{T}_{ab} = & \frac{18 \pi \ell^4 \chi^2(c^2 r^4-12 \pi^2 \ell^8 \chi^2)}{c r^2(c^2 r^4 + 36 \pi^2 \ell^8 \chi^2)} q_{ab} + \left( -3 + \frac{4 c^2 r^4}{c^2 r^4 + 36 \pi^2 \ell^8 \chi^2} \right)T_{ab}\, ,
    \end{align}
\end{subequations}
with
\begin{equation}
    \mathcal{T} = \frac{12 \pi \ell^6 c\, \chi^2 }{c^2 r^4 + 36 \pi^2 \ell^8 \chi^2}\, , \qquad \Ottbar = - \frac{2c^2 \ell^4 \chi^2}{c^2 r^4 + 36 \pi^2 \ell^8 \chi^2}\, .
\end{equation}
The boundary action for this solution is the same as \eqref{bdry-action-non-cov-bc-null} and the corresponding symmetry algebra is the same as the extended CSS geometry, $u(1)$-Kac-Moody algebra extended by a Heisenberg pair, see \ref{appen:charge-section5.3} for the analysis. 
\section{Boundary conditions inside \texorpdfstring{AdS$_3$}{AdS3}} \label{sec:bc-finite}
In the previous two sections we constructed and analyzed solutions associated with various boundary conditions imposed at asymptotic AdS$_3$ boundary. Freelance holography setting, however, also allows considering any arbitrary timelike codimension-one boundary. In this section, we showcase solutions associated with Dirichlet boundary conditions on $\Sigma_c=\Sigma_{r=r_c}$, in section \ref{sec:D-bc-finite}. Then, we work through a similar construction for conformal and Neumann  boundary conditions in sections \ref{sec:C-bc-finite} and \ref{sec:N-bc-finite}. 

\subsection{Dirichlet inside \texorpdfstring{AdS$_3$}{AdS3}} \label{sec:D-bc-finite}
We first consider the Dirichlet condition on $\Sigma_c=\Sigma_{r=r_c}$, which can also be interpreted as a specific form of mixed boundary condition at the asymptotic boundary $\Sigma$. We explore this using two complementary approaches.

\paragraph{First approach, solving Einstein's equation.}
The Dirichlet boundary condition on the finite-radius hypersurface $\Sigma_c$ is given by
\begin{equation}
    \delta h_{ab}(r_c,x^a) = 0\, .
\end{equation}
As can be seen from \eqref{h-coord-trans}, this condition can be interpreted as a particular mixed boundary condition for the asymptotic data $\{q_{ab}, T^{ab}\}$ at the boundary $\Sigma$. As in the Ba\~nados case, we fix $h_{ab}(r_c,x^a)$ as follows
\begin{equation}
    \d{}s_{h}^2 = h_{ab}(r_c, x^a) \d{}x^{a} \d{}x^{b} = - r_c^2 \d{}x^{+} \d{}x^{-}\, .
\end{equation}
Next, we should solve the following momentum and Hamiltonian constraints
\begin{equation}
    {\nabla}_b{\mathcal{T}}^{b}_{a}=0\, , \qquad  \mathcal{T} + \frac{6\pi\ell^2}{{c}}\Ottbar  = 0\, .
\end{equation}
The momentum constraint leads to the following expressions
\begin{equation}\label{rBY-EBT-D-rc}
   \begin{split}
        & \mathcal{T}_{\pm\pm} = L_\pm(x^\pm) + \partial^2_{\pm}X(x^+,x^-)\, , \\
        & \mathcal{T}_{+-} = - \partial_{+}\partial_{-}X(x^+,x^-)\, .
   \end{split}
\end{equation}
To determine $X(x^+,x^-)$, we must apply the Hamiltonian constraint, which takes the form
\begin{equation}
    \partial_{+}\partial_{-} X + \frac{12 \pi \ell^2}{ c r_c^2} \Biggl[  ({L}_+ + \partial_{+}^2 X)  ({L}_- + \partial_{-}^2 X) - (\partial_{+}\partial_{-} X)^2 \Biggr] = 0\, .
\end{equation}

Solving the equation above determines $X$ as a functional of ${L}_{\pm}$. To proceed, we solve it perturbatively in the small parameter $\mu$ and Taylor-expand $X$ as a power series in $\mu$
\begin{equation}
    X= \sum_{n=1}^{\infty} \frac{\mu^n}{n!} X_n\, ,\qquad \mu:=\frac{12\pi\ell^2}{ c r_c^2},
\end{equation}
Note that $\mu$ becomes small at large $r_c$.  Substituting this expansion into the equation and solving order by order yields
\begin{equation}
    \begin{split}
      &  X_1 = - \mathcal{L}_- \mathcal{L}_+\, , \\
      & X_2 = \mathcal{L}_+^2 \partial_{-}\mathcal{L}_- + \mathcal{L}_-^2 \partial_{+}\mathcal{L}_+\, , \\
    \end{split}
\end{equation}
where $\mathcal{L}_{\pm}(x^\pm)$ are related to $L_{\pm}(x^\pm)$ as
\begin{equation}\label{anti-deriv}
     {\cal L}_{\pm} (x^{\pm})= \int^{x^\pm} L_\pm(x^\pm)\, ,\qquad L_\pm(x^\pm)=\partial_\pm {\cal L}_{\pm} (x^{\pm})\, .
\end{equation}
In general, the coefficients $X_n$ take the form
\begin{equation}\begin{split}
    X_n = (-1)^n \sum_{p=1}^n \frac{1}{p} \binom{n}{p-1}\, \partial_+^{n-p} \mathcal{L}_+^p\,  \partial_-^{p-1} \mathcal{L}_-^{n-p+1}  \, .
\end{split}
\end{equation}
As a result, the full expression for $X$ can be written as a double sum
\begin{equation}\label{X-exp-D}
X=\sum_{q=1}^\infty \sum_{p=1}^\infty\ \frac{(-\mu)^{p+q-1}}{p! q!} \partial_+^{q-1} \partial_-^{p-1}\ ({\cal L}_+^p {\cal L}_-^q)\, .
\end{equation}
This expression is reminiscent of the double-sum structures for dressed operators introduced in \cite{Chen:2025jzb} from a purely field-theoretic perspective. Here, however, we derive them directly from the gravity side.

\paragraph{Second approach, coordinate transformation.}
We now turn to an alternative approach based on \textit{field-dependent} coordinate transformations, following the method outlined in \cite{Guica:2019nzm}. As mentioned earlier, in $3d$ Einstein gravity, starting from a flat AdS boundary and moving inward into the bulk results in constant-radius hypersurfaces that also have flat induced metrics. This is a distinctive feature of $3d$ gravity, and that in $2d$ (on any constant $r$ surface) Einstein tensor vanishes identically. Consequently, the induced metric at any finite radial slice can be related to that at the asymptotic boundary through a suitable coordinate transformation. Here, we employ such transformations to implement the Dirichlet boundary condition at a finite radial location. 

We begin with the induced metric of the Ba\~nados geometry on $\Sigma_c$, given in \eqref{Banados-3D}
\begin{equation}
        \d{}s_{h}^2 = - \left( r_c \d{}x^+ - \ell^2 \frac{L_-}{r_c} \d{}x^-\right)\left( r_c \d{}x^- - \ell^2 \frac{L_+}{r_c} \d{}x^+\right)\, .
\end{equation}
Clearly, the induced metric on $\Sigma_c$ does not satisfy the Dirichlet boundary condition $\delta h_{ab}\big|_{r_c} \neq 0$. Our goal is to perform a field-dependent coordinate transformation such that the metric takes a simpler form. To this end, we consider the following condition:
\begin{equation}
     \left(\d{}x^+ -\frac{ \ell^2 }{r_c^2}L_- \d{}x^-\right)\left( \d{}x^- - \frac{\ell^2 }{r_c^2} L_+\d{}x^+\right) =  \d{}X^{+} \d{}X^{-}\, .
\end{equation}
A coordinate transformation that satisfies this relation is
\begin{equation}\label{coordinate-trans}
    X^{\pm}(x^+,x^-) = x^{\pm} -\frac{ \ell^2 }{r_c^2} \mathcal{L}_{\mp}(x^\mp)\, ,
\end{equation}
where $\mathcal{L}_{\mp}$ and $L_\pm$ are related as $L_\pm(x^\pm)=\partial_\pm {\cal L}_{\pm} (x^{\pm})$.  
A crucial point here is that the coordinate transformation above is field-dependent; when applied to a given solution, it modifies the boundary conditions accordingly. Recalling the analysis of subsection~\ref{sec:CPSF}, any modification of boundary conditions is generated by a $W$-term, which is holographically interpreted as a  multi-trace deformation of the holographic boundary theory. Taken together, these facts imply that a field-dependent coordinate transformation can itself be understood as a multi-trace deformation of the boundary theory.\footnote{That field-dependent coordinate transformations yield  changes in  boundary conditions was noted in \cite{Afshar:2017okz, Grumiller:2019ygj}.}

The next step is to express the Ba\~nados line element \eqref{Banados-3D} in terms of the new coordinates $X^{\pm}$. For this, we compute
\begin{equation}
    \d{}x^{\pm} = \frac{\d{}X^\pm + \frac{\ell^2}{r_c^2}L_\mp \d{}X^{\mp}}{1-\frac{\ell^4}{r_c^4}L_{+} L_-}.
\end{equation}
Therefore, the $3d$ line element \eqref{Banados-3D} with Dirichlet boundary condition on $\Sigma_c$ takes the form
\begin{equation}
    \begin{split}
        \d{}s^{2} = \ell^2\frac{\d{} r^2}{r^2} - \frac{r^2}{1-\frac{\ell^4}{r_c^4}L_{+}L_-} &\left[ \big( 1 - \frac{\ell^4}{r^2 r_c^2}L_{+}L_{-} \big)\d{}X^{+} +\big(\frac{\ell^2}{r_c^2}- \frac{\ell^2}{r^2} \big) L_- \d{}X^{-} \right] \times \\
        &\left[ \big( 1 - \frac{\ell^4}{r^2 r_c^2}L_{+}L_{-} \big)\d{}X^{-} + \big(\frac{\ell^2}{r_c^2}- \frac{\ell^2}{r^2} \big) L_+\d{}X^{+} \right]\, .
    \end{split}
\end{equation}
In this expression, the functions $L_{\pm}$ are understood as functions of $X^{\pm}$, i.e., $L_{\pm}(x^{\pm}(X^+,X^-))$. To make this dependence explicit, one needs to invert the transformation \eqref{coordinate-trans}. Manifestly, at $r=r_c$ the above line element reduces to $-r_{c}^2 \d{}X^{+}\d{}X^{-}$. In the  $r_c\gg \ell$ limit, it reproduces the Ba\~nados family \eqref{Banados-3D}, as expected. To compare this metric with the Ba\~nados family (with Dirichlet boundary conditions at large $r$), we note that the conformal induced metric at the AdS boundary is given by
\begin{equation}
    \d{}s_{q}^{2} =  - \frac{\ell^2}{1-\frac{\ell^4}{r_c^4}L_{+}L_-} \left( \d{}X^{+} +\frac{\ell^2}{r_c^2}L_-\d{}X^{-} \right) \left( \d{}X^{-} +\frac{\ell^2}{r_c^2}L_+  \d{}X^{+} \right)\, .
\end{equation}
This explicitly demonstrates that the metric at the AdS boundary fluctuates and is no longer compatible with a Dirichlet boundary condition; instead, it gives rise to a mixed boundary condition.

We also use the relation between the transformed and original stress tensors at large $r$,
\begin{equation}\label{EMT-infty-F-D-bc}
\begin{split}
    {\mathfrak{T}}_{ab} \d{}X^{a} & \d{}X^b := T_{ab} \d{}x^{a} \d{}x^{b}\, \\
    & = - \frac{\frac{c}{12\pi}}{(1-\frac{\ell^4}{r_c^4}L_{+} L_-)^2} \Bigg[ 4\frac{\ell^2}{r_c^2}L_+ L_- \d{}X^+ \d{}X^{-} 
       +\left(1+\frac{\ell^4}{r_c^4}L_+ L_-\right)\Big(L_+ (\d{}X^{+})^2+L_- (\d{}X^{-})^2\Big)\Bigg].
\end{split}
\end{equation}
Then the trace of ${\mathfrak{T}}_{ab} $ is given by
\begin{equation}
     {\mathfrak{T}} =  \frac{4 \ell^2c}{3\pi r_c^4}\frac{L_+ L_-}{(1-\frac{\ell^4}{r_c^4}L_{+} L_-)^2}\, .
\end{equation}
Finally, we write the EMT at a finite distance
\begin{equation}
\begin{split}
    \mathcal{T}_{ab} \d{}X^{a}\, \d{}X^{b} &=  - \frac{c r_{c}^4\, L_{+}}{12\pi r^2}\frac{r^{6}r_{c}^4+\ell^4 L_{+}L_{-}[r^{6}-6 r^{4}r_{c}^2+3r^2 r_c^4 + (3r^2-2r_c^2)\ell^4 L_{+}L_{-}]}{(r^4 - \ell^4 L_{+}L_{-})(r_c^4-\ell^4 L_+ L_-)^2} (\d{}X^{+})^{2} \\
    & \hspace{-.16 cm} - \frac{c r_{c}^4\, L_{-}}{12\pi r^2}\frac{r^{6}r_{c}^4+\ell^4 L_{+}L_{-}[r^{6}-6 r^{4}r_{c}^2+3r^2 r_c^4 + (3r^2-2r_c^2)\ell^4 L_{+}L_{-}]}{(r^4 - \ell^4 L_{+}L_{-})(r_c^4-\ell^4 L_+ L_-)^2} (\d{}X^{-})^{2} \\
    & \hspace{-.16 cm}+\frac{c \ell^2 r_c^4 L_+ L_-}{6\pi r^2} \frac{r^{4}r_{c}^2(3r_c^2-2r^2)+\ell^{4}L_{+}L_{-}(3r^4-6r^2 r_c^2+r_c^4+\ell^4 L_+L_-)}{(r^4-\ell^4 L_+ L_-)(r_c^4-\ell^4 L_+ L_-)^2}\d{}X^{+}\d{}X^{-}.
\end{split}
\end{equation}
One can readily verify that the asymptotic boundary EMT \eqref{EMT-infty-F-D-bc} is recovered in the $r \to \infty$ limit.

{Finally, we note that the asymptotic symmetries, their associated surface charges, and the corresponding algebras for AdS$_3$ with Dirichlet boundary conditions at a finite cutoff were studied in \cite{Kraus:2021cwf}, where a one-parameter nonlinear deformation of the standard Virasoro algebra was obtained. Moreover, the boundary action governing the dynamics of the boundary modes was derived in \cite{Ebert:2022cle} and shown to be described by a Nambu–Goto–type action.}

\subsection{Conformal inside \texorpdfstring{AdS$_3$}{AdS3}} \label{sec:C-bc-finite}
To construct solutions with a conformal boundary condition imposed at a finite distance, we need to solve the constraint at a given $r=r_c$ surface
\begin{equation}\label{const-conf-finite-r}
  \mathcal{T} + \frac{c}{24\pi} \mathcal{R} + \frac{6\pi\ell^2}{c}\, \Ottbar = 0 \, , 
  \qquad \nabla_b \mathcal{T}^{b}{}_{a}=0 \, ,
\end{equation}  
together with the boundary conditions \eqref{NC-bc-def}
\begin{equation}
   \delta\!\left(\frac{h_{ab}}{\sqrt{-h}}\right)=0 \, , \qquad \delta \mathcal{T}=0 \, .
\end{equation}
For simplicity, we restrict to a flat boundary $\mathcal{R}=0$ and adopt the explicit form  
\begin{equation}
   \mathcal{T} = \mathcal{T}_0 \, , \qquad \frac{h_{ab}}{\sqrt{-h}} = -2\, \d{}x^{+} \d{}x^{-} \, ,
\end{equation}  
where $\mathcal{T}_0$ is a constant with zero variation, $\delta \mathcal{T}_0 =0$. For this case, \eqref{const-conf-finite-r} implies that $\delta \Ottbar=0$ and hence $\Ottbar=-\frac{c\, \mathcal{T}_0}{6\pi \ell^2}$. The solution is then given by at $r=r_c$
\begin{equation}
   h_{ab}\, \d{}x^a \d{}x^b = - h_{+}(x^+) h_{-}(x^-)\, \d{}x^+ \d{}x^- \, , 
\end{equation}
and 
\begin{equation}\label{conformal-inside}
    \mathcal{T}_{ab} \d{}x^a \d{}x^b = \frac{A}{4} h_{+}^2  (\d{}x^+)^2 - \frac{\mathcal{T}_0}{2} h_+ h_- \d{}x^+ \d{}x^- +  \frac{B}{4} h_-^2 (\d{}x^-)^2\, ,\qquad  AB= \mathcal{T}_0 (\mathcal{T}_0- \frac{c}{3 \pi \ell^2})\,.
\end{equation}

A few remarks are in order:  
\begin{enumerate}
\item The metric in \eqref{conformal-inside} may be compared to \eqref{calT-conformal}. Using boosts which scale $x^\pm \to \gamma^{\pm 1} x^\pm$, one can always set $A=B$, as in \eqref{calT-conformal}. Therefore, one learns that 
$\mathcal{T}_0(\mathcal{T}_0- \frac{c}{3 \pi \ell^2})\geq 0$.

\item The induced metric and the corresponding rBY-EMT are independent of $r$, and hence remain valid at any radial position. The line element of the complete three-dimensional spacetime is given by
    \begin{equation}
        \d{}s^2 = \ell^2 \frac{\d{}r^2}{r^2}  -  r^2 h_{+}(x^+) h_{-}(x^-)\, \d{}x^+ \d{}x^-\, .
    \end{equation}
    \item Unlike the Dirichlet boundary condition at finite distance, which leads to the non-local solution \eqref{rBY-EBT-D-rc} with \eqref{X-exp-D}, the conformal boundary condition yields a local result.  
\end{enumerate}
\subsection{Neumann inside \texorpdfstring{AdS$_3$}{AdS3}} \label{sec:N-bc-finite}
The case of Neumann boundary conditions at a finite radial distance requires no new construction: as shown in subsection~\ref{sec:N-bc-infty}, the Neumann boundary condition does not flow. In other words, imposing Neumann at infinity guarantees that it remains Neumann at any finite radius. More explicitly, the solution with Neumann boundary conditions at radius $r$ is given by equations~\eqref{metric-N-bc-Sigma} and~\eqref{EMT-N-r}. Interestingly, in contrast to the Dirichlet case at finite distance, the Neumann boundary condition continues to yield a local solution.
\section{Outlook}\label{sec:discussion}
Freelance holography \cite{Parvizi:2025shq, Parvizi:2025wsg} provides a framework that generalizes standard holography in two key directions: it accommodates arbitrary boundary conditions for bulk fields and allows holography to be formulated on any timelike hypersurface. In this work, we explored freelance holography in three dimensions in detail. In particular, we derived the deformation flow equation for the boundary action on an arbitrary timelike boundary with a generic boundary condition. Owing to the special features of three-dimensional gravity, we were able to solve these flow equations exactly. This generalizes the previous construction of finite-cutoff holography from the case of Dirichlet boundary conditions \cite{Guica:2019nzm, Taylor:2018xcy, Hartman:2018tkw} to arbitrary ones. Furthermore, we examined how different boundary conditions deform as the boundary flows from the asymptotic AdS boundary to a finite radial position in the bulk.

Our construction provides the framework to systematically find classically consistent boundary conditions on an arbitrary timelike boundary. To show this at work, we explicitly constructed solution spaces for different boundary conditions at the asymptotic boundary of AdS. Besides reanalyzing previously known and studied examples of  Dirichlet boundary conditions (Ba\~nados geometries) \cite{Banados:1998gg, Compere:2015knw, Sheikh-Jabbari:2016unm}, conformal boundary conditions \cite{Afshar:2017okz, Adami:2020uwd}, CSS geometries  \cite{Compere:2013bya} and black flowers \cite{Afshar:2016kjj, Afshar:2016wfy} within our framework, we also constructed and analyzed new solutions with Neumann boundary conditions and cases with ``non-covariant'' boundary conditions. Of course, our explicit solution space construction is not exhaustive, while the framework is general. 

\paragraph{Field-dependent coordinate transformations, boundary conditions, and change of slicing.} As pointed out in \cite{Parvizi:2025shq, Parvizi:2025wsg} and reviewed in section \ref{sec:Basic-setup}, the change of boundary conditions is associated with the change of the $W$-term (boundary Lagrangian). Another closely related concept is ``change of slicing'' \cite{Adami:2020ugu, Adami:2021nnf, Geiller:2021vpg, Adami:2021sko, Adami:2022ktn} over the solution space. The change of slicings are generated by \textit{field-dependent} coordinate transformations. One can prove that \cite{In-progress-change-of-slicing} $W$-terms are generators of the change of slicing. Therefore, one would expect that changes in boundary conditions and slicings should be physically related, if not identical. In particular, one should be able to show that Dirichlet, Neumann, black flower, and CSS solutions can be obtained from the conformal case via field-dependent coordinate transformations. Examples of such coordinate transformations have been discussed in \cite{Afshar:2017okz}. Moreover, one can consider $r$-dependent, field-dependent coordinate transformations to relate different boundary conditions at different radii. In such cases, see examples in \cite{Grumiller:2019ygj, Adami:2020ugu}, one can interpolate between different boundary conditions at different radii. It would be instructive to develop a freelance holography program by exploring more systematically the role of field and $r$ -dependent diffeomorphisms. 

As our analysis in appendix \ref{appen:charges} shows, the symmetry algebras of the class of solutions corresponding to different boundary conditions are ``symplectic'' \cite{Compere:2015knw} as we are dealing with a solution phase space. The symmetry algebras depend on the boundary conditions; however, all of them fall into the class of algebras discussed in \cite{FarahmandParsa:2018ojt}, which may typically be obtained as a deformation of BMS$_3$ or $u(1)$-Kac-Moody algebras. The notion of deformation of algebras and the change of slicing (and hence changes of boundary conditions) are closely related to each other. It can be instructive to explore this relation in more detail. 

\paragraph{Freelance fluid/gravity correspondence.}
It is known that gauge/gravity correspondence admits the limit fluid/gravity correspondence where (conformal) relativistic hydrodynamics of the boundary field theory side is well described by Einstein’s field equations in the bulk  \cite{Bhattacharyya:2007vjd, Rangamani:2009xk, Hubeny:2011hd}. One can extend the freelance holography program and construct a freelance fluid/gravity correspondence. Using the tools of freelance holography, one can interpolate between the fluid description at a finite cutoff surface and at infinity, thereby constructing the dictionary that relates finite-distance fluid variables to their asymptotic counterparts. Moreover, one can study how the fluid/gravity correspondence should be modified as we change the boundary conditions.

In the $3d$ (bulk) case,  the absence of bulk propagating modes is expected to allow one to establish an exact fluid/gravity correspondence and describe the boundary fluid as a perfect fluid \cite{In-progress-freelance-hydro}. In higher dimensions, however, the presence of bulk modes induces dissipative effects \cite{Bhattacharyya:2007vjd, Rangamani:2009xk, Hubeny:2011hd} (see also \cite{Arenas-Henriquez:2025rpt} for a recent discussion on the connection between bulk modes and dissipation). A natural direction for future work is to study how deformations of the asymptotic fluid variables—particularly the dissipative transport coefficients—flow as the boundary is moved inward, thereby providing an interpolation between the asymptotic coefficients and their finite-distance counterparts.

\paragraph{Detailed analysis of higher-dimensional freelance holography.}
In this work, we focused on freelance holography in three-dimensional Einstein gravity, where the absence of bulk propagating modes allows for an exact integration of Einstein’s equations. In higher dimensions, however, such an exact treatment is no longer possible. A perturbative analysis remains feasible: one can solve the radial dependence of the metric perturbatively near the AdS boundary \cite{Skenderis:2002wp, deHaro:2000vlm}, and thereby construct finite-cutoff holography at large but finite distances. Within this framework, the flow of boundary conditions can also be systematically developed. 

\paragraph{Boundary induced gravity.}
In \cite{Adami:2025pqr}, an intriguing observation was made: the boundary T$\bar{\text{T}}$ deformation can be equivalently interpreted as the emergence of boundary gravity. The analysis, however, was restricted to the AdS$_5$/CFT$_4$ correspondence. Extending this perspective to three-dimensional gravity and searching for a gravitational interpretation of the T$\bar{\text{T}}$ deformation in that setting is another promising direction.
\paragraph{Freelance holography behind the horizon.}
Finite-cutoff holography has so far been largely restricted to regions outside the black hole horizon. Recent work \cite{AliAhmad:2025kki} has extended this framework to probe the interior, in particular, the vicinity of the black hole singularity. While that analysis focused on Dirichlet holography, it is interesting to generalize it within the framework of freelance holography, allowing for arbitrary boundary conditions.
\paragraph{Effective field theory approach.}
Recently, \cite{Allameh:2025gsa} studied finite-cutoff holography on AdS$_3$ with conformal boundary conditions and proposed that the boundary theory can be understood as a deformation of the standard asymptotic QFT, supplemented by a timelike Liouville theory that accounts for the dynamics of new scalar degrees of freedom introduced by the conformal boundary condition, and further deformed by a dressed T$\bar{\text{T}}$ operator \footnote{{See also \cite{Galante:2025tnt} for the setup with a conformal boundary placed near the AdS boundary, where the effective boundary dynamics is captured by a one-dimensional Liouville or by two coupled Schwarzian theories.}}. This construction provides a concrete example within the framework of freelance holography. Extending this type of analysis to other boundary conditions and computing physical observables on both sides of the duality to test the proposal presents another intriguing direction for future work.

{
\paragraph{Addition of matter fields.}
In this paper, we focused on pure AdS$_3$ gravity and analyzed the RG flow of various boundary conditions. An interesting direction for future work is to include additional fields (see \cite{Arkhipova:2024iem} for a discussion of holographic RG flows and boundary conditions in three-dimensional gauged supergravity). This extension would allow one to study deformations beyond the T$\bar{\text{T}}$ case \cite{Smirnov:2016lqw, Guica:2017lia, Conti:2019dxg, LeFloch:2019rut} and to explore the boundary condition flows associated with other fields.}

\begin{acknowledgments}
We would like to thank Hamed Adami for his collaboration on related topics and for his contributions during the early stages of this project. We thank Ali Parvizi for collaboration on similar subjects. MMShJ gratefully acknowledges the hospitality of the Beijing Institute of Mathematical Sciences and Applications (BIMSA), where this project was initiated. The work of VT is supported by the Iran National Science Foundation (INSF) under project No. 4040771.
\end{acknowledgments}

\appendix

\section{Some useful relations}\label{appen:useful-rel}
Having integrated the $r$-dependence of Einstein's equations, one can work out useful identities relating finite arbitrary $r$ quantities to those at asymptotic large $r$:
\begin{subequations}\label{h-q--q-h}
\begin{align}
h_{ab} =\left(\frac{r}{\ell}\right)^2 {\cal Q}_{ac} q^{cd}{\cal Q}_{db}\, ,&\qquad  {\cal Q}_{ab}:=q_{ab} - \frac{6\pi\ell^4}{c r^2}\tilde{T}_{ab}\, , 
\label{hab-again}\\
q_{ab} = \left(\frac{\ell}{r}\right)^2 {\cal H}_{ac} h^{cd}{\cal H}_{db}\, ,&\qquad  {\cal H}_{ab}:=h_{ab} +\frac{6\pi\ell^2}{c }\tilde{{\cal T}}_{ab}\, , 
\label{qab-again}
\end{align}\end{subequations}
\begin{subequations}\label{All-r-quantities-1}
\begin{align}
({\cal Q}^{-1})^{ab}=  \left(\frac{r}{\ell}\right)^2 \frac{\sqrt{-q}}{\sqrt{-h}}\ \big(q^{ab}+\frac{6\pi\ell^4}{cr^2} T^{ab}\big)\,, &\qquad ({\cal H}^{-1})^{ab}= \left(\frac{\ell}{r}\right)^2 \frac{\sqrt{-h}}{\sqrt{-q}}\  \big(h^{ab}-\frac{6\pi\ell^2}{c} {\cal T}^{ab}\big)\,, \label{Q-H-inverse}\\
{\cal Q}_{ab}=\left(\frac{\ell}{r}\right)^2 {\cal H}_{ab}\, ,&\qquad ({\cal Q}^{-1})^{ab}= \left(\frac{r}{\ell}\right)^2 ({\cal H}^{-1})^{ab}\,,\label{Q-H}\\
       \tilde{\mathcal{T}}_{ab}  = {\cal Q}_{ac} q^{cd} \tilde{T}_{db} \, ,&\qquad 
     \tilde{T}_{ab} =  {\cal H}_{ac} h^{cd} \tilde{\mathcal{T}}_{db}\,, \label{T-calT}\\
     \sqrt{-h}\ {\mathcal{T}}^a{}_{b}&=\left(\frac{r}{\ell}\right)^2 {\sqrt{-q}}\ {T}^a{}_{b}\, , \label{hcalT-qT} \\
       \sqrt{-h}\, \mathcal{T}^{ab} & = \left(\frac{\ell}{r}\right)^2 \sqrt{-q} (\mathcal{Q}^{-1})^{ac}q_{cd} T^{db}\, ,
\end{align}\end{subequations}
\begin{subequations}\label{All-r-quantities-2}
\begin{align}
   \sqrt{-h}\, \mathcal{T} = \sqrt{-q} \big(T-\frac{6\pi \ell^4}{r^2c} \ttbarb\big)\, ,& \qquad 
   \sqrt{-q}\, {T} = \sqrt{-h} \big({\cal T}+\frac{6\pi \ell^2}{c} \Ottbar\big)\, ,\label{qT-hcalT}\\ 
    \sqrt{-h}\, {\cal R} =\sqrt{-q}\, {R}\, &,\qquad
    \sqrt{-h}\, \Ottbar = \frac{\ell^2}{r^2} \sqrt{-q}\ \ttbarb\,, \label{r-dep-ttbar-3}\\
      \sqrt{-h} = \sqrt{-q} \left( \frac{r^2}{\ell^2} -\frac{\ell^2}{4} {R}  -\frac{18 \pi^2 \ell^6 \ttbarb}{c^2 r^2}\right)\, &,\qquad    \sqrt{-q} = \frac{\ell^2}{r^2} \sqrt{-h} \left( 1+ \frac{\ell^2}{4} {\cal R}  +\frac{18 \pi^2 \ell^4}{c^2 }\Ottbar\right)\, ,\label{detq-deth} 
\end{align}\end{subequations}
where $\ttbarb$ is the T$\bar{\text{T}}$ operator associated with boundary EMT
\begin{equation}
    \ttbarb := T^{ab} T_{ab} - T^2\, .
\end{equation}
In deriving the above equations, we have used the following $2d$ identities
\begin{equation}
    \begin{split}
        T_{a}^{c} T_{cb} = T T_{ab} + \frac{1}{2} \ttbarb q_{ab}\, ,\qquad \tilde{T}_{a}^{c} \tilde{T}_{cb} = -T \tilde{T}_{ab} + \frac{1}{2} \ttbarb q_{ab}\, , \qquad {\tilde T}_{a}^{c} T_{cb} = \frac{1}{2} \ttbarb q_{ab}\, .
    \end{split}
\end{equation}
From the above and that $\sqrt{-q}, \sqrt{-h}$ are positive and do not change sign, we learn that $\Ottbar$ and $\ttbarb$ have the same sign and 
\begin{equation}\label{T-TTbar-ineq}
    {\cal T}+\frac{3\pi\ell^2}{c}\Ottbar \leq \frac{c}{6\pi \ell^2}.
\end{equation}
For flat boundary case with $R=0$,  $T=0$ and ${\cal R}=0$, and thus $\Ottbar \geq -\frac{c^2}{18\pi^2 \ell^4}$ and ${\cal T} \leq \frac{c}{3\pi \ell^2}$.

\section{Examples of geometries with non-flat boundary} \label{appen:non-flat-bc}
The families of solutions discussed in the main text all have the property that the constant $r$ slices are flat geometries. In this appendix, we briefly discuss cases where the $2d$ boundary metric is not flat. Having the boundary metric $q_{ab}$ and energy momentum tensor $T_{ab}$, it is then straightforward to use the equations worked out in section \ref{sec:Basic-setup} to read out the complete $3d$ solution.

\paragraph{Dirichlet boundary conditions}\label{non-flat-D-bc}
In principle, the $2d$ boundary metric $q_{ab}$ or $h_{ab}$ need not be flat. We know that any $2d$ metric can be written as conformally flat
\begin{equation}
    \d{}s^2 = 2 e^{\phi} \d{}x^+ \d{}x^{-}\, ,
\end{equation}
where $\phi=\phi(x^+,x^-)$. For a given $\phi$ this metric satisfies Dirichlet boundary condition at infinity. Now we solve the constraint equations
\begin{equation}
    \begin{split}
        & T_{\pm \pm} =L_{\pm}(x^{\pm}) +\frac{c}{48\pi} \left[ (\partial_{\pm}\phi)^2 - 2 \partial_{\pm}^2 \phi \right]\, , \\
        &T_{+-} = \frac{c}{24 \pi} \partial_{+}\partial_{-} \phi\, .
    \end{split}
\end{equation}
It is clear that the variational principle (i.e., the symplectic potential) cannot vanish unless $\delta\phi = 0$, meaning that $\phi$ is fixed and does not vary over the solution space. For this case 
\begin{equation}\begin{split}
R=-\frac{24\pi}{c} &T= -2 e^{-\phi} \partial_{+}\partial_{-} \phi\, , \qquad \sqrt{-q}= e^{\phi}\, , \\ \ttbarb=2 &e^{-2\phi}\left(T_{++}T_{--}- (T_{+-})^2\right)\, .
\end{split}\end{equation}
Let us look at the cases where the AdS boundary has a constant curvature. Then, we need to require 
\begin{equation}
    -2 e^{-\phi} \partial_{+}\partial_{-} \phi = \bar{R}\, ,
\end{equation}
where $ \bar{R}$ is a constant. We aim to solve this equation for $\phi$; it is the Liouville equation
\begin{equation}\label{2d-curvature}
    \partial_{+}\partial_{-} \phi + \frac{\bar{R}}{2}\, e^{\phi} = 0\, .
\end{equation}
Its solution is given by
\begin{equation}
   \begin{split}
      & \bar{R} = 0: \quad \phi(x^+,x^-) =  f+g\, , \\
      & \bar{R} \neq 0: \quad \phi(x^+,x^-) = \ln \left[ \frac{4 \partial_+ f \partial_{-} g}{\bar{R} (1+f g)^2} \right]\, , 
    \end{split}
\end{equation}
where $f=f(x^+)$ and $g=g(x^-)$ are two arbitrary given functions of their arguments. 

With a given (with vanishing variation) $\phi$, the above describes solutions with Dirichlet boundary conditions. If $\bar{R}$ is positive (negative), these are slicings of locally AdS$_3$ geometries by dS$_2$ (AdS$_2$) geometries. 

\paragraph{Neumann boundary conditions}

As the next solution to \eqref{2d-curvature}, consider the following line element for the AdS boundary
\begin{equation}
    \d{}s_{q}^{2} =\frac{8 f' h'}{\bar{R} (1+f h)} (\d{}t \d{}\phi +\d{}\phi^2)\, ,
\end{equation}
where $h(\phi)$ and $f(t+\phi)$ are two arbitrary functions. The corresponding EMT is also given by
\begin{equation}
    \sqrt{-q}\, T^{ab} = \frac{c \bar{R}}{48 \pi } \begin{bmatrix}
2 & -1 \\
-1 & 0
\end{bmatrix}\, ,
\end{equation}
which clearly satisfies the Neumann boundary condition at infinity. For this case, we have
\begin{equation}
   T = - \frac{c \bar{R}}{ 24 \pi }\, , \qquad \ttbarb = \frac{1}{2}\left( \frac{c \bar{R}}{24 \pi} \right)^{2}\, .
\end{equation}

\paragraph{Conformal boundary conditions}
We note that the solution constructed in subsection \ref{non-flat-D-bc} can also be interpreted as a solution with a conformal boundary condition:
\begin{equation}
\frac{q_{ab}}{\sqrt{-q}}  \mathrm{d}x^{a} \mathrm{d}x^{b} = 2 \mathrm{d}x^{+} \mathrm{d}x^{-} \,, \qquad T = - \frac{c \bar{R}}{24 \pi} \,.
\end{equation}
Furthermore, since $\delta \phi = 0$, it follows that $\delta \bar{R} = 0$.

\section{Surface charge analysis}\label{appen:charges}
In this appendix, we analyze the symmetries and corresponding surface charges of the solutions under various boundary conditions discussed in the main text. We also compute the associated symmetry and surface charge algebras. This analysis allows us to determine whether the arbitrary functions appearing in the solutions correspond to physical degrees of freedom or are pure gauge. All the symmetry algebras obtained here are symplectic symmetries of the solution phase spaces and fall into the classes of algebras analyzed in \cite{FarahmandParsa:2018ojt}.
\subsection{Neumann soft hair}\label{appen:Neumann-charges}
In this subsection, we focus on the Neumann case \eqref{metric-N-bc-Sigma}.
\paragraph{Symmetry generators.} The symmetries that preserve the form of the metric \eqref{metric-N-bc-Sigma} are given by
\begin{equation}\label{sym-N}
    \xi^{\mu}\partial_{\mu} = \Big[-2 t T + \frac{\ell^2(\ell^2 - t r^2 h)}{2r^2 f}\partial_{\phi}T + t \Big(\partial_{\phi}Y- \frac{\ell^2}{f} \partial_{\phi}^2 T \Big) \Big]\partial_t + r T \partial_{r} + Y \partial_{\phi}\, ,
\end{equation}
where \( T(\phi) \) and \( Y(\phi) \) denote arbitrary symmetry generators corresponding to supertranslations and superrotations, respectively. We have also introduced \( h = \frac{g - 2\,\partial_{\phi} f}{f} \).
\paragraph{Symmetry algebra.} Under the adjusted bracket \cite{Barnich:2010eb, Compere:2013bya}, the algebra of these symmetry generators is given by
\begin{equation}
    [\xi(T_1, Y_1), \xi(T_2, Y_2)]_{\text{\tiny{adj}}} = \xi(T_{12}, Y_{12})\, ,
\end{equation}
with 
\begin{equation}
    T_{12} = Y_{1} \partial_{\phi} T_{2} - Y_{2} \partial_{\phi}T_1\, , \qquad Y_{12} = Y_{1} \partial_{\phi} Y_{2} - Y_{2} \partial_{\phi} Y_1\, .
\end{equation}
This is the semidirect sum of the Witt algebra and an abelian current algebra.
\paragraph{Surface charge variation.} The Lee-Wald surface charge \cite{Lee:1990nz} (see \cite{Grumiller:2022qhx} for a review) associated with symmetry \eqref{sym-N} is also given by
\begin{equation}\label{charge-N}
    \delta Q_\xi =  \int \d{}\phi \left( T\, \delta L + Y\, \delta J \right)\, , \quad \text{with} \quad J := -\frac{\ell}{16 \pi G}\, h\, , \quad L := \frac{1}{8 \pi G\ell} f\, . 
\end{equation}
This result shows that the functions \( f(\phi) \) and \( g(\phi) \) appearing in \eqref{metric-N-bc-Sigma} parametrize non-trivial diffeomorphisms, as they give rise to non-vanishing surface charges. The charge aspects transform as follows
\begin{equation}
   \begin{split}
       & \delta_\xi J = \partial_{\phi} (Y J) + \frac{\ell}{4\pi G} \partial_{\phi}T + \frac{\ell}{8\pi G} \partial_{\phi}^2 Y\, , \\
       & \delta_\xi L = J \partial_{\phi}T + Y \partial_\phi L + 2 L \partial_\phi Y -\frac{1}{8\pi G\ell} \partial_{\phi}^2 T\, .
   \end{split}
\end{equation}
\paragraph{Charge algebra.} 
The charge algebra of \eqref{charge-N} is also given by
\begin{equation}
    \{Q_{\xi_1}, Q_{\xi_2} \} = Q_{[\xi_1,\xi_2]_{\text{\tiny{adj}}}} + K_{\xi_1,\xi_2}\, ,
\end{equation}
where $ K_{\xi_1,\xi_2}$ is the central extension term of the algebra
\begin{equation}
    K_{\xi_1,\xi_2} = \frac{\ell}{8\pi G}\int \d{}\phi (T_1 \partial_{\phi}T_2-T_2 \partial_{\phi}T_1 + T_1 \partial_{\phi}^2 Y_2 - T_2 \partial_{\phi}^2 Y_1)\, .
\end{equation}
The explicit form of the algebra for the Fourier modes of $L, J$ charges, respectively $L_n, J_n$, is as follows
\begin{equation}\label{charge-algebra-Neumann}
   \begin{split}
        &[ L_m, L_n]= (m-n) L_{m+n} \\ 
        & [L_m, J_n]= n J_{m+n} + \frac{\ell}{4G} \delta_{m+n,0} \\ 
        & [J_m, J_n]= \frac{\ell}{4G}\ m \delta_{m+n,0}, 
   \end{split}
\end{equation}
which is a $u(1)$ Kac–Moody algebra with two central extensions: one in the commutator of the two $u(1)$ currents (the level of Kac-Moody) and the other in the commutator of the Witt generators with the current. Note that this is different from the usual $u(1)$ Kac–Moody algebra, e.g., the one obtained in the CSS case \cite{Compere:2013bya}, where the central extensions are in the Virasoro (Witt) part and the level of the $u(1)$ current.\footnote{We note that the $u(1)$ Kac-Moody in general admits three different central extensions; e.g., see \cite{FarahmandParsa:2018ojt}.}
\subsection{Conformal soft hair}\label{appen:conformal-charges}
In this subsection, we consider the surface charges associated with the conformal solution \eqref{conf-soln}.
\paragraph{Symmetry generator.} The vector fields preserving the form of the metric \eqref{conf-soln} are given by
\begin{equation}\label{sym-C}
    \xi^\mu \partial_\mu = \xi^+(x^+)\partial_+ + \xi^-(x^-) \partial_-\, ,
\end{equation}
where $\xi^+(x^+)$ and $\xi^-(x^-)$ are two arbitrary functions of their arguments and parametrize supertranslations in the null directions $x^\pm$.
\paragraph{Symmetry algebra.} The algebra of symmetries \eqref{sym-C} is written as follows
\begin{equation}
    [\xi(\xi^+_1, \xi^-_1), \xi(\xi^+_2, \xi^-_2)]_{\text{\tiny{adj}}} = \xi(\xi^+_{12}, \xi^-_{12})\, ,
\end{equation}
with 
\begin{equation}
    \xi^+_{12} = \xi^+_1 \partial_+ \xi^+_2 - \xi^+_2 \partial_+ \xi^+_1\, , \qquad \xi^-_{12} = \xi^-_1 \partial_- \xi^-_2 - \xi^-_2 \partial_- \xi^-_1 \, .
\end{equation}
This is two copies of the Witt algebra.
\paragraph{Surface charge variation.}
The corresponding surface charge variation is
\begin{equation}
    \delta Q_\xi = \int \d{}\phi (\xi^+ \delta L_+ + \xi^- \delta L_-)\, , \quad \text{with} \quad L_+ := \frac{h_+^2}{16\pi G \ell} \, , \quad L_-:= \frac{h_-^2}{16\pi G \ell}\, .
\end{equation}
Transformations of charge aspects under symmetries \eqref{sym-C} are as follows
\begin{equation}
    \begin{split}
    & \delta_\xi L_{\pm} = \xi^{\pm} \partial_{\pm} L_{\pm} + 2 L_\pm \partial_\pm \xi^\pm \, .
    \end{split}
\end{equation}
\paragraph{Charge algebra.}
The charge algebra is the same as the algebra of symmetry generators (two copies of the Witt algebra) \textit{without} any central extension term. We comment that, upon a change of slicing, one could have taken $\zeta_\pm=\xi_\pm/h_\pm$ as the new field-independent symmetry generators. In this case one would have obtained two $u(1)$ current algebras (instead of Witt) \cite{Afshar:2017okz, Adami:2020uwd}.
\subsection{Soft hair of black flowers}\label{appen:blackflower-charges}
In this subsection, we compute the symmetries and the surface charge variations of black flowers \eqref{blackflower-metric}. As the previous cases, here too, we deal with symplectic symmetries. 
\paragraph{Symmetry generator.} Symmetries of this solution are generated by
\begin{equation}\label{sym-BF}
    \xi^{\mu} \partial_{\mu} = T(\phi) \partial_t + Y(\phi) \partial_{\phi}\, ,
\end{equation}
where $T(\phi)$ and $Y(\phi)$ parametrize supertranslations and superrotations respectively.
\paragraph{Symmetry algebra.}
These symmetry generators satisfy the following algebra
\begin{equation}
    [\xi(T_1, Y_1), \xi(T_2, Y_2)]_{\text{\tiny{adj}}} = \xi(T_{12}, Y_{12})\, ,
\end{equation}
with 
\begin{equation}
    T_{12} = Y_{1} \partial_{\phi} T_{2} - Y_{2} \partial_{\phi}T_1\, , \qquad Y_{12} = Y_{1} \partial_{\phi} Y_{2} - Y_{2} \partial_{\phi} Y_1\, .
\end{equation}
This is the semidirect sum of the Witt algebra and an abelian current algebra.
\paragraph{Surface charge variation.}
The surface charge expression associated with these symmetries is as follows
\begin{equation}
    \delta Q_\xi = \int \d{}\phi (T \delta M + Y \delta J)\, , \quad \text{with} \quad M :=\frac{1}{32\pi G} J_+\, , \quad J := - \frac{ \ell}{32\pi G} J_+ J_-\, .
\end{equation}
The charge aspects of the black flower solution transform under the symmetries \eqref{sym-BF} as follows
\begin{equation}
    \begin{split}
        \delta_\xi M = \partial_{\phi}(Y\, M) \,, \qquad \delta_\xi J = M \partial_{\phi} T + Y \partial_{\phi} J + 2 J \partial_{\phi}Y \, .
    \end{split}
\end{equation}
\paragraph{Charge algebra.} The charge algebra coincides with the algebra of symmetry generators, \textit{without} any central extension term.
\subsection{CSS soft hair}\label{appen:CSS-charges}
In this subsection, we analyze the symmetries and surface charges of the CSS solution \eqref{CSS-soln}.
\paragraph{Symmetry generators.} 
The vector fields that preserve the form of the metric \eqref{CSS-soln} are given by
\begin{equation}
    \begin{split}
        \xi^\mu \partial_\mu = & \left[T + \frac{1}{4}\left( 2t W +\frac{c\ell^4(-c r^2 + 3\pi \ell^4 \Delta J) \partial_{\phi}^2 Y}{c^2 r^4 - 9 \pi^2 \ell^8 \Delta (3\Delta J^2 + 4L)} \right) \right]\partial_t - \frac{r}{4}(W+2 \partial_{\phi}Y) \partial_r \\
        &+ \left[ Y -\frac{3c \pi \ell^8 \Delta \partial_{\phi}^2 Y}{2c^2 r^4 - 18 \pi^2 \ell^8 \Delta (3\Delta J^2 + 4L)} \right] \partial_{\phi}\, ,
    \end{split}
\end{equation}
where \( T(\phi) \) and \( Y(\phi) \) are arbitrary functions parameterizing the supertranslations and superrotations, respectively, and \( W \) is an arbitrary constant.
\paragraph{Symmetry algebra.}
Next, we consider the symmetry algebra, which yields
\begin{equation}
    [\xi(T_1, Y_1, W_1), \xi(T_2, Y_2, W_2)]_{\text{\tiny{adj}}} = \xi(T_{12}, Y_{12}, W_{12})\, ,
\end{equation}
with 
\begin{equation}
  W_{12} = 0\, , \qquad  T_{12} = Y_{1} \partial_{\phi} T_{2} +\frac{1}{2}W_2 T_1 - (1\xleftrightarrow{}2)\, , \qquad Y_{12} = Y_{1} \partial_{\phi} Y_{2} - (1\xleftrightarrow{}2)\, .
\end{equation}
\paragraph{Surface charge variation.}
The charge variation is then given by
\begin{equation}
    \slashed{\delta} Q_\xi =  \int \d{}\phi \left[ Y \delta \Big(L + \frac{\Delta}{2} J^2 \Big)  - \sqrt{\Delta}\, T\, \delta(\sqrt{\Delta} J) \right] - \frac{ t}{2}\sqrt{\Delta} W \delta (\sqrt{\Delta}\, J_0)\, .
\end{equation}
This expression suggests defining new generators
\begin{equation}
    \hat{W}= \sqrt{\Delta} W\, , \qquad \hat{T} = \sqrt{\Delta} T\, , \qquad \hat{Y} = Y\, ,
\end{equation}
and accordingly, new charge aspects 
\begin{equation}
    \mathcal{L} := L + \frac{\Delta}{2} J^2 \, , \qquad  \mathcal{J} := - \sqrt{\Delta}\, J\, .
\end{equation}
In terms of these new generators and new variables, the charge variation becomes integrable\footnote{This technique, which renders charge variations integrable by introducing new generators and corresponding charge aspects, is known as the "change of slicing" method \cite{Adami:2020ugu, Adami:2021nnf, Adami:2022ktn, Taghiloo:2024ewx, Ruzziconi:2020wrb, Geiller:2021vpg}.
}
\begin{equation}
    \delta Q_\xi =  \frac{\pi\, t}{2} \hat{W} \delta \mathcal{J}_0+ \int \d{}\phi \left( \hat{Y} \delta \mathcal{L}  + \hat{T}\, \delta \mathcal{J} \right) \, ,
\end{equation}
where $2\pi \mathcal{J}_0 \equiv \int_0^{2\pi} \d{}\phi \mathcal{J}(\phi)$.
The variation of these new fields takes the following form
\begin{equation}
    \begin{split}
        & \delta_\xi \mathcal{J} = \partial_{\phi} (\hat{Y} \mathcal{J}) - 2 \partial_{\phi} \hat{T}\, ,\\
        & \delta_\xi \mathcal{L} = \mathcal{J} \partial_{\phi} \hat{T} + \hat{Y} \partial_{\phi} \mathcal{L} + 2 \mathcal{L} \partial_{\phi} \hat{Y} + \frac{c}{24\pi} \partial_{\phi}^3 \hat{Y}\, ,\\
        &\delta_\xi \sqrt{\Delta}= \frac12 \hat{W}\, .
    \end{split}
\end{equation}
The above also shows that $\delta_\xi  \mathcal{J}_0=0$, i.e. $\mathcal{J}_0$ is the center of the soft charge algebra.  
\paragraph{Symmetry algebra in new slicing.}
The symmetry algebra, expressed in terms of these new generators, takes the following form
\begin{equation}
    [\xi(\hat{T}_1, \hat{Y}_1, \hat{W}_1), \xi(\hat{T}_2, \hat{Y}_2, \hat{W}_2)]_{\text{\tiny{adj}}} = \xi(\hat{T}_{12}, \hat{Y}_{12}, \hat{W}_{12})\, ,
\end{equation}
with 
\begin{equation}
  \hat{W}_{12} = 0\, , \qquad  \hat{T}_{12} = 
  \hat{Y}_{1} \partial_{\phi} \hat{T}_{2} - (1\xleftrightarrow{}2)\, , \qquad \hat{Y}_{12} = \hat{Y}_{1} \partial_{\phi} \hat{Y}_{2} - (1\xleftrightarrow{}2)\, .
\end{equation}
\paragraph{Charge algebra.}
The CSS charge algebra is given by
\begin{equation}
    \{Q_{\xi_1}, Q_{\xi_2} \} = Q_{[\xi_1,\xi_2]_{\text{\tiny{adj}}}} +  K_{\xi_1,\xi_2} \, ,
\end{equation}
with the following central extension term
\begin{equation}
     K_{\xi_1,\xi_2} = \frac{c}{48\pi}\int \d{}\phi \left[ \hat{Y}_1 \partial_{\phi}^3 \hat{Y}_2 -  \hat{Y}_1 \partial_{\phi}^3 \hat{Y}_2 + \frac{48 \pi}{c} (\hat{T}_2 \partial_{\phi} \hat{T}_1 - \hat{T}_1 \partial_{\phi} \hat{T}_2)\right]\, .
\end{equation}
The explicit form of the algebra can be read as
\begin{equation}\label{u1KM-Heisenberg}
   \begin{split}
        &[ L_m, L_n]= (m-n) L_{m+n}+\frac{\ell}{8G} {n^3} \delta_{m+n,0}\, , \\ 
        & [L_m, J_n]= - n J_{m+n}\, , \\ 
        & [J_m, J_n]= -2 m \delta_{m+n,0}\, , 
        \\  &{ [\mathcal{J}_0, \sqrt{\Delta}]= \frac{i}{2\pi} }\,,
   \end{split}
\end{equation}
which is a $u(1)$ Kac–Moody algebra with two central extensions plus a center piece, the former is precisely the one in \cite{Compere:2013bya} and the center piece has appeared due to the extension of the CSS case we have considered (cf. discussions in section \ref{sec:CSS}).
\subsection{Soft hair for the solution phase space in section \ref{sec:new-c-bc}}\label{appen:charge-section5.2}
In this subsection, we present an analysis of the symmetries and conserved charges related to subsection \ref{sec:new-c-bc}.
\paragraph{Symmetry generators.} The following symmetries preserve the form of the three-dimensional line element described in subsection \ref{sec:new-c-bc}
\begin{equation}
    \begin{split}
        & \xi^t = t W + T + \frac{\tilde{\Delta}}{\Delta} Y - \frac{ c(c r^2 + 6\pi \ell^4 \tilde{\Delta})(c r^2 \partial_{t}^2 T - 6\pi \ell^4 \Delta\, \partial_{\phi}^2 Y)}{24 \pi \Delta (36 \pi^2 \ell^8 \Delta\, g - c^2 r^4 f)} \, , \\
        & \xi^r = - \frac{r}{2} ( \tilde{W} - \partial_t T + \partial_\phi Y) \, , \\
        & \xi^\phi = Y - \frac{c \ell^4 (c r^2 \partial_{t}^2 T- 6\pi \ell^4 \Delta \partial_\phi^2 Y)}{4(36 \pi^2 \ell^8 \Delta\, g - c^2 r^4 f)}\, ,
    \end{split}
\end{equation}
where \( Y = Y(\phi) \) and \( T = T\!\left(t - \frac{\tilde{\Delta}}{\Delta}\phi\right) \) are arbitrary functions of their respective arguments, parameterizing the superrotations and supertranslations, respectively. In addition, there are two constant symmetry generators, \( W \) and \( \tilde{W} \).
\paragraph{Symmetry algebra.}
Next, we examine the corresponding symmetry algebra, which yields
\begin{equation}
    [\xi(T_1, Y_1, W_1, \tilde{W}_1), \xi(T_2, Y_2, W_2, \tilde{W}_1)]_{\text{\tiny{adj}}} = \xi(T_{12}, Y_{12}, W_{12}, \tilde{W}_{12})\, ,
\end{equation}
with 
\begin{equation}
  \begin{split}
      & W_{12} = 0\, , \quad \tilde{W}_{12}=0\, , \quad T_{12} = T_{1} \partial_{t} T_{2} + T_1 W_2 - (1\xleftrightarrow{}2)\, , \quad Y_{12} = Y_{1} \partial_{\phi} Y_{2} - (1\xleftrightarrow{}2)\, .
  \end{split}
\end{equation}
\paragraph{Surface charge variation.}
The charge variation is given by
\begin{equation}\label{charge-1-5.2}
    \slashed{\delta} Q_\xi =  \int \d{}\phi \Big[ Y \delta g  + \frac{\Delta}{\tilde{\Delta}} T\, \delta \Big( \frac{\tilde{\Delta}^2}{\Delta} f\Big) \Big] + t \frac{\Delta}{\tilde{\Delta}} W  \delta \Big( \frac{\tilde{\Delta}^2}{\Delta} f_0\Big)\, ,
\end{equation}
where $f_0 \equiv \int f \d{}\phi$ is the zero mode of $f$. The above expression suggests the following change of slicing
\begin{equation}
    \mathcal{F} = \frac{\tilde{\Delta}^2}{\Delta} f\, , \qquad \mathcal{F}_0 \equiv \int \d{}\phi\, \mathcal{F}\, ,
\end{equation}
\begin{equation}
    \hat{T} = \frac{\Delta}{\tilde{\Delta}} T\, , \qquad \hat{W} =  \frac{\Delta}{\tilde{\Delta}} W\, , \qquad \hat{Y} = Y\, .
\end{equation}
Expressed in these new variables, the surface charge variation becomes
\begin{equation}
   {\delta} Q_\xi =  \int \d{}\phi ( Y \delta g  + \hat{T}\, \delta \mathcal{F} ) + t \hat{W}  \delta \mathcal{F}_0\, .
\end{equation}
The charge aspects transform as follows
\begin{equation}\label{charge-aspec-trans-5.2}
    \begin{split}
        & \delta_\xi g = Y \partial_\phi g + 2g \partial_\phi Y + \frac{c}{24\pi} \partial_{\phi}^3 Y\, , \\
        & \delta_\xi \mathcal{F} = - \hat{T} \partial_{\phi} \mathcal{F} -2 \mathcal{F} \partial_{\phi} \hat{T} - \frac{c}{24\pi} \partial_{\phi}^3 \hat{T}\, . \\
        & \delta_\xi (\frac{\Delta}{\tilde\Delta}) = \hat{W}\, , \qquad \delta_\xi \tilde{\Delta} = \tilde{\Delta}\, \tilde{W}\, .
    \end{split}
\end{equation}
\paragraph{Symmetry algebra in new slicing.}
The algebra of hatted generators is
\begin{equation}
    [\xi(\hat{T}_1, \hat{Y}_1, \hat{W}_1, \hat{\tilde{W}}_1), \xi(\hat{T}_2, \hat{Y}_2, \hat{W}_2, \hat{\tilde{W}}_1)]_{\text{\tiny{adj}}} = \xi(\hat{T}_{12}, \hat{Y}_{12}, \hat{W}_{12}, \hat{\tilde{W}}_{12})\, ,
\end{equation}
with 
\begin{equation}
  \begin{split}
      & \hat{W}_{12} = 0\, , \quad \hat{\tilde{W}}_{12}=0\, , \quad \hat{T}_{12} = \hat{T}_{2} \partial_{\phi} \hat{T}_{1} - (1\xleftrightarrow{}2)\, , \quad \hat{Y}_{12} = \hat{Y}_{1} \partial_{\phi} \hat{Y}_{2} - (1\xleftrightarrow{}2)\, .
  \end{split}
\end{equation}  
This shows algebra of $\hat{T}$ and $\hat{Y}$ are two copies of Witt algebras.
\paragraph{Charge algebra.}
The charge algebra of this solution space is given as follows
\begin{equation}
    \{Q_{\xi_1}, Q_{\xi_2} \} = Q_{[\xi_1,\xi_2]_{\text{\tiny{adj}}}} +  K_{\xi_1,\xi_2} \, ,
\end{equation}
with the following central extension terms
\begin{equation}
     K_{\xi_1,\xi_2} = \frac{c}{48\pi}\int \d{}\phi \left( \hat{Y}_{1} \partial_{\phi}^3 \hat{Y}_2 - \hat{Y}_2 \partial_{\phi}^3 \hat{Y}_1 - \hat{T}_1 \partial_{\phi}^3 \hat{T}_2 +  \hat{T}_2 \partial_{\phi}^3 \hat{T}_1 \right)\, .
\end{equation}
This shows $g(\phi)$ and $\mathcal{F}(t- \frac{\tilde{\Delta}}{\Delta}\phi)$ construct two copies of the Virasoro algebras at Brown-Henneaux central charge. We also note that zero modes satisfies 
\begin{equation}\label{F0Delta}
    [\mathcal{F}_0,\frac{\Delta}{\tilde{\Delta}}]=\frac{i}{2}\, .
\end{equation}
Finally, we emphasize that only a specific combination of $\Delta$ and $\tilde{\Delta}$ appears in the charge variation \eqref{charge-1-5.2}. Consequently, only one of them is physical, while the other is a pure gauge and can be fixed. From the last line of \eqref{charge-aspec-trans-5.2}, we can choose $\delta \tilde{\Delta} = 0$, which then implies $\tilde{W} = 0$, for which choice \eqref{F0Delta} may be written as $[{\cal F}_0, \Delta]=\frac{i\tilde\Delta}{2}$.
\subsection{Soft hair for the solution phase space in section \ref{sec:new-non-covariant}}\label{appen:charge-section5.3}
\paragraph{Symmetry generators.} Diffeomorphisms which move us within the family of solutions \eqref{sec:new-non-covariant} are
\begin{equation}\label{sym-5.3}
    \xi^\mu \partial_\mu = \left[T(\phi) + \frac{t}{2}W \right] \partial_t - \frac{r}{2}W \partial_r + Y(\phi) \partial_\phi\, ,
\end{equation}
whose adjusted Lie brackets are
\begin{equation}
    [\xi(T_1, Y_1, W_1), \xi(T_2, Y_2, W_2)]_{\text{\tiny{adj}}} = \xi(T_{12}, Y_{12}, W_{12})\, ,
\end{equation}
with 
\begin{equation}
  W_{12} = 0\, , \qquad  T_{12} = Y_{1} \partial_{\phi} T_{2} +\frac{1}{2}W_2 T_1 - (1\xleftrightarrow{}2)\, , \qquad Y_{12} = Y_{1} \partial_{\phi} Y_{2} - (1\xleftrightarrow{}2)\, .
\end{equation}
\paragraph{Surface charge variation.}
The Lee-Wald surface charge variation associated with the symmetries \eqref{sym-5.3} is given by
\begin{equation}
    \slashed{\delta} Q_\xi =  \int \d{}\phi \Big\{ Y \delta \big[ \chi(f^2 + g^2) \big]  + \sqrt{\chi}\, T\, \delta(2\sqrt{\chi} g) \Big\} + \frac{t}{2}\sqrt{\chi} W \delta (2\sqrt{\chi}\, g_0)\, .
\end{equation}
where $g_0 := \int_0^{2\pi} \d{}\phi g(\phi)$. With the following re-namings, 
\begin{equation}
    \hat{W}= \sqrt{\chi}\,  W\, , \qquad \hat{T} = \sqrt{\chi}\, T\, , \qquad \hat{Y} = Y\, ,
\end{equation}
\begin{equation}
   \mathcal{L} := \chi (f^2 + g^2) \, , \qquad  \mathcal{J} := 2 \sqrt{\chi}\, g\, .
\end{equation}
\begin{equation}
    \delta Q_\xi =  \frac{\pi\, t}{2} \hat{W} \delta \mathcal{J}_0+ \int \d{}\phi \left( \hat{Y} \delta \mathcal{L}  + \hat{T}\, \delta \mathcal{J}\right) \, ,
\end{equation}
where $2\pi \mathcal{J}_0 := \int_0^{2\pi} \d{}\phi\, \mathcal{J}(\phi)$. The charge aspects transform as follows
\begin{equation}
    \begin{split}
        & \delta_\xi \mathcal{J} = \partial_{\phi} (\hat{Y} \mathcal{J}) + 2 \partial_{\phi} \hat{T}\, ,\\
        & \delta_\xi \mathcal{L} = \mathcal{J} \partial_{\phi} \hat{T} + \hat{Y} \partial_{\phi} \mathcal{L} + 2 \mathcal{L} \partial_{\phi} \hat{Y}\, ,\\
        &\delta_\xi \sqrt{\chi}= \frac12 \hat{W}\, .
    \end{split}
\end{equation}
{Algebra of the hatted symmetry generators takes the form}
\begin{equation}
    [\xi(\hat{T}_1, \hat{Y}_1, \hat{W}_1), \xi(\hat{T}_2, \hat{Y}_2, \hat{W}_2)]_{\text{\tiny{adj}}} = \xi(\hat{T}_{12}, \hat{Y}_{12}, \hat{W}_{12})\, ,
\end{equation}
with 
\begin{equation}
  \hat{W}_{12} = 0\, , \qquad  \hat{T}_{12} = 
  \hat{Y}_{1} \partial_{\phi} \hat{T}_{2} - (1\xleftrightarrow{}2)\, , \qquad \hat{Y}_{12} = \hat{Y}_{1} \partial_{\phi} \hat{Y}_{2} - (1\xleftrightarrow{}2)\, .
\end{equation}
\paragraph{Charge algebra.}
The charge algebra takes the form
\begin{equation}
    \{Q_{\xi_1}, Q_{\xi_2} \} = Q_{[\xi_1,\xi_2]_{\text{\tiny{adj}}}} +  K_{\xi_1,\xi_2} \, ,
\end{equation}
with
\begin{equation}
     K_{\xi_1,\xi_2} = \int d\phi \left( \hat{T}_1 \partial_{\phi} \hat{T}_2 - \hat{T}_2 \partial_{\phi} \hat{T}_1 \right)\, .
\end{equation}
This yields a $u(1)$ Kac-Moody algebra, analogous to \eqref{u1KM-Heisenberg}, with vanishing central charge in the Virasoro sector and with $\Delta$ replaced by $\chi$.
\bibliographystyle{fullsort.bst}
\bibliography{reference}
\end{document}